\documentclass[preprint,notoc]{JHEP3}
\usepackage{epsfig}

\def\to{\rightarrow}

\def\bi{\begin{itemize}}
\def\ei{\end{itemize}}
\def\te{\tilde e}

\def\ttau{\tilde \tau}

\def\tell{\tilde\ell}
\def\tq{\tilde q}
\def\tw{\widetilde W}
\def\tz{\widetilde Z}

\def\be{\begin{equation}}  
\def\ee{\end{equation}}  

\title{Direct Detection of Dark Matter\\ in Supersymmetric Models}

\author{Howard Baer, Csaba Bal\'azs, Alexander Belyaev\footnote
{On leave of absence from Nuclear Physics Institute, Moscow State University.}~
and Jorge O'Farrill 
\\ Department of Physics, Florida State University Tallahassee, FL 32306, USA\\
E-mail: \email{baer@hep.fsu.edu}, \email{balazs@hep.fsu.edu},
        \email{belyaev@hep.fsu.edu}, \email{ofarrill@hep.fsu.edu}}

\preprint{\vbox{\hbox{FSU-HEP-030515}}} 

\abstract{
We evaluate neutralino-nucleon scattering rates in several well-motivated 
supersymmetric models, and compare against constraints on the neutralino relic 
density, $BF( b\to s\gamma )$ as well as the muon anomalous magnetic moment 
$a_\mu$. In the mSUGRA model, the indirect constraints favor the hyperbolic 
branch/focus point (HB/FP) region of parameter space, 
and in fact this region is 
just where neutralino-nucleon scattering rates 
are high enough to be detected in 
direct dark matter search experiments! In Yukawa unified SUSY $SO(10)$ models 
with scalar mass non-universality, the relic density of neutralinos is almost 
always above experimental bounds, 
while the corresponding direct detection rates 
are below experimental levels. Conversely, in five dimensional $SO(10)$ models 
where gauge symmetry breaking is the result of compactification of the extra 
dimension, and supersymmetry breaking is communicated via gaugino mediation, the 
relic density is quite low, while direct detection rates can be substantial.
}


\begin{document}

\section{Introduction}
\label{sec:intro}

A wide variety of astrophysical data now point conclusively to the existence of 
cold dark matter (CDM) in the universe. The most recent results come from the 
Wilkinson Microwave Anisotropy Probe~(WMAP)\cite{wmap}. Their results confirm 
the standard model of cosmology and fit its parameters to high precision. The 
properties of a flat universe in the $\Lambda CDM$ model are characterized by 
the density of baryons ($\Omega_b$), matter density ($\Omega_m$), vacuum energy 
($\Omega_\Lambda$) and the expansion rate ($h$) which are measured to be:
\begin{eqnarray}
\Omega_b&=&0.044\pm0.004 \\
\Omega_m&=&0.27\pm0.04   \\
\Omega_\Lambda&=&0.73\pm0.04 \\
h&=&0.71^{+0.04}_{-0.03} .
\end{eqnarray}
From the WMAP results one derives the following value for the cold dark matter 
density:
\begin{equation}
\Omega_{CDM}h^2=0.1126^{+0.0081}_{-0.0090}
( {^{+0.0161}_{-0.0181}} )  \mbox{ at 68(95)\% CL}.
\end{equation}

In spite of the excellent fit to the data, the standard cosmological model has 
several fundamental open questions.  One of these is the origin and nature of 
dark matter.  Within the context of $R$-parity conserving supersymmetry 
(SUSY)~\cite{reviews} the lightest supersymmetric particle (LSP) offers a robust 
solution to this problem.
First, SUSY itself is the most complete theoretical extension of the Standard 
Model (SM), and is well motivated experimentally.  This solidly supports 
the neutralino LSP dark matter theoretically.
Second, the WMAP results strongly restrict neutrino hot dark matter to 
$\Omega_\nu h^2<0.0076\ (95\% CL.)$ and also exclude warm dark matter based on 
detection of re-ionization at $z\sim 20$.  Warm dark matter in the form of 
gravitinos could also be a good candidate for dark matter in certain 
supersymmetric models, but WMAP has vetoed this option.  
Therefore, in the light of recent WMAP results, the neutralino LSP emerges as 
one of the most attractive candidate for CDM.

If indeed all space is filled with relic neutralinos, then it may be possible to 
directly detect them via their scattering from nuclei, or to indirectly detect 
them via their annihilation products. 
There are two kinds of non-accelerator experiments aimed at the CDM search.  One 
way is to search for the products of neutralino annihilation in outer space or 
inside the sun or earth where they become concentrated due to the gravitational 
force.  Another way to search for CDM is via direct detection from their 
scattering off nuclei by measuring the nuclear recoil. 
There are several existing and future projects engaged in the direct search for 
weakly interacting massive particles (WIMPS)~\cite{wimp-det-reviews}.  These 
include experiments based on germanium detectors such as IGEX~\cite{igex}, 
HDMS~\cite{hdms}, CDMS~\cite{CDMS}, EDELWEISS~\cite{EDELWEISS} and
GENIUS~\cite{GENIUS}.  Scintillator detectors are used by DAMA~\cite{DAMA} and
ZEPLIN~\cite{ZEPLIN1,ZEPLIN2,ZEPLIN3,ZEPLIN4}.  Projection chambers utilized in 
DRIFT~\cite{spooner}, and metastable particle detectors in SIMPLE~\cite{simple} 
and PICASSO~\cite{picasso}. 

Presently the best
limits on the spin-independent WIMP-nucleon cross-section ($\sigma_{SI}$) have 
been obtained by the CDMS, EDELWEISS and ZEPLIN1 groups, while a signal is 
claimed by the DAMA collaboration. 
Collectively, we will refer to the reach from these groups as the ``Stage 1''
dark matter search. Depending on the neutralino mass,
the combined limit on the neutralino-proton spin-independent cross section 
$\sigma_{SI}$
varies from  $10^{-5}$ to $10^{-6}$~pb. This cross section
range is beyond the predicted levels from most supersymmetric models if one 
also requires the model to be within the 
experimental constraints from LEP2, $(g-2)_\mu$, $BF( b\to s\gamma )$, 
relic density $\Omega_{CDM}h^2$
and $BF( B_s\to \mu^+\mu^- )$ data. 
However, experiments in the near future
like CDMS2, CRESST2, ZEPLIN2 and EDELWEISS2 (Stage 2 detectors) should have a 
reach of the order of 
$10^{-8}$~pb.
Finally, a number of experiments such as GENIUS, ZEPLIN4 and CRYOARRAY are in 
the planning stage. We refer to them as Stage 3 detectors, which promise 
strong limits of the order of $\sigma_{SI}<10^{-9}$ -- $10^{-10}$~pb, and 
would allow the exploration of a considerable part of parameter space of many 
supersymmetric models.

Beginning with the paper by Goodman and Witten~\cite{goodman}, there have been 
numerous studies on neutralino nuclei scattering rate 
evaluation~\cite{griest,drees-nojiri,baer-brhlik,dirdet-ellis,dirdet,dirdet-nu}, 
with the trend of gradually improving the quality of calculations and extending 
the range of supersymmetric models.  It is useful to note that in the case of 
the general MSSM, $\sigma_{SI}$ could be several orders of magnitude 
higher~\cite{dirdet-nu} than in more constrained models such as the minimal 
supergravity model (mSUGRA).

In this paper, we perform an updated analysis of neutralino elastic scattering 
off nuclei in a class of supersymmetric models.  There are various motivations 
for this study.
\begin{itemize}
\item
We have included constraints from the neutralino relic density 
$\Omega_{\tilde{Z}_1} h^2$, the rare decays $b\to s \gamma$ and 
$B_s\to\mu^+\mu^-$, the muon anomalous magnetic moment $a_\mu$, as well as from 
the LEP2 experiment. Together these constraints significantly restrict the SUSY 
model parameter space in the relevant range of $\sigma_{SI}$.
\item
We are using a new version of ISAJET, version 7.65\cite{isajet}, in which the 
complete one-loop corrections to sfermion masses~\cite{bagger-corr} have been 
incorporated.  In addition, the stability of the new version has been greatly 
improved in the low $|\mu|$ region which allows us to better access the 
hyperbolic branch/focus point (HB/FP) region of the mSUGRA model. This is 
important, because the HB/FP region can simultaneously satisfy constraints from 
$\Omega_{\tilde{Z}_1} h^2$, $(g-2)_\mu$, and $BF(b\to s\gamma )$ while offering 
a decoupling solution to the SUSY flavor and CP problems, and maintaining 
naturalness\cite{ccn,fmm,sug_chi2}. 
\item
Besides the mSUGRA model, we also analyse SUSY grand unified theories (GUTs) 
based on the $SO(10)$ group~\cite{so10-susyguts} where the MSSM Higgs doublets 
are both present in the same 10-dimensional Higgs multiplet.  These models 
predict unification of the Yukawa couplings for the third generation: 
$f_t=f_b=f_\tau$.  
Following the strategy of our previous study~\cite{so10-4}, we examined models 
with a high degree of Yukawa coupling unification, and have evaluated neutralino 
nuclei scattering rates together with experimental constraints.  For Yukawa 
unified models with $\mu <0$, there exist parameter choices consistent with 
neutralino relic density constraints, although direct detection rates are 
frequently quite low.  For models with $\mu >0$, the relic density is almost 
always above measured limits, while direct detection rates are almost always 
below experimentally accessible levels.
\item
We also examine an $SO(10)$ motivated SUSY GUT model \cite{pati-salam} with 
non-universal gaugino masses and otherwise vanishing soft-parameters at the GUT 
scale. This setup is well motivated by SUSY GUTs constructed in higher 
dimensions~\cite{5d+su5,5d+so10} utilizing gaugino mediated SUSY breaking. Many 
of these models elegantly solve the doublet-triplet splitting and fast proton 
decay problems of 4D SUSYGUT models~\cite{4d-prob} by the oribifold 
compactification of the extra dimenstions. 
In the phenomenologically viable region of the parameter space of this model the 
lightest neutralino generally has a large higgsino or wino component, which 
leads to low values of $\Omega_{\tz_1}h^2$, and neutralino dark matter would 
need to be augmented by other forms of CDM, such as axions, or hidden sector 
states. We find that $\sigma_{SI}$ can frequently be several orders of magnitude 
larger than the mSUGRA case, owing to the large wino and/or higgsino component 
of the neutralino. 

\end{itemize}

The rest of this paper is organized as follows.  In Section~2 we discuss details 
of our evaluation  of the  neutralino-nucleon elastic scattering cross section. 
In Section~3, we present results for the mSUGRA model, and point out the 
intriguing features endemic to the HB/FP region.  In Section~4, we perform 
studies for $SO(10)$ motivated Yukawa unified SUSY GUT models, while Section~5 
presents results for SUSY GUT models with non-universal gaugino masses.  In 
Section~6 we present our conclusions.

\section{Details of calculation for neutralino elastic scattering on nuclei}

The interactions for elastic scattering of neutralinos on nuclei can be
described by the sum of spin-independent (${\cal L}^{eff}_{scalar}$) and 
spin-dependent (${\cal L}^{eff}_{spin}$) Lagrangian terms:
\begin{equation}
 {\cal L}^{eff}_{elastic}={\cal L}^{eff}_{scalar}+{\cal L}^{eff}_{spin} .
\end{equation}
In this paper we evaluate the spin-independent cross section $\sigma_{SI}$ of 
neutralino scattering off of nuclei which is the main experimental observable 
since $\sigma_{SI}$ contributions from individual nucleons in the nucleus add 
coherently and can be expressed via SI nuclear form-factors. The cross section 
$\sigma_{SI}$ receives contributions from neutralino-quark interactions via 
squark, $Z$ and Higgs boson exchanges, and from neutralino-gluon interactions 
involving quarks, squarks and Higgs bosons at the 1-loop level. The differential 
$\sigma_{SI}$ off a nucleus $X_Z^A$ with mass $m_A$ takes the 
form~\cite{kam_review}
\begin{equation}
 \frac{d\sigma^{SI}}{d|\vec{q}|^2}=\frac{1}{\pi v^2}[Z f_p +(A-Z) f_n]^2 
 F^2 (Q_r),
\end{equation}                                 
where $\vec{q}=\frac{m_A m_{\widetilde Z_1}}{m_A+m_{\widetilde Z_1}}\vec{v}$ is 
the three-momentum transfer, $Q_r=\frac{|\vec{q}|^2}{2m_A}$ and $F^2(Q_r)$ is 
the scalar nuclear form factor, $\vec{v}$ is the velocity of the incident 
neutralino and $f_p$ and $f_n$ are effective neutralino couplings to protons and 
neutrons respectively. 
This formalism has been reviewed in~\cite{kam_review,drees-nojiri,baer-brhlik}. 
Explicit expressions for $f_p$ and $f_n$ can be found, {\it e.g.} 
in~\cite{baer-brhlik}. 
The original calculation has been done in~\cite{drees-nojiri} and can 
be expressed as
\begin{equation}
{f_N \over m_N} = \sum_{q=u,d,s} \frac{f_{Tq}^{(N)}}{m_q} \left[
f_q^{(\tq )}+f_q^{(H)}-{1\over 2}m_q m_{\tz_1}g_q\right]  + 
\frac{2}{27} f^{(N)}_{TG} \sum_{c,b,t} \frac{f_q^{(H)}}{m_q} +\cdots
\end{equation}
where $N=p,\ n$ for neutron, proton respectively, and
$f^{(N)}_{TG} = 1 - \sum_{q=u,d,s} f^{(N)}_{Tq} $.
The expressions for the $f_q^{(H)}$ couplings as well as other terms
denoted by $\cdots$ are omitted for the sake of brevity but can be found 
in~\cite{drees-nojiri,baer-brhlik}.

The parameters  $f_{Tq}^{(p)}$, defined by
\begin{equation}
<N|m_q \bar{q} q|N> = m_N f_{Tq}^{(N)} ~~~~~ (q=u,d,s)
\end{equation}
contains uncertainties due to errors on the experimental measurements of quark masses.
We have adopted values of renormalization-invariant  constants $f_{Tq}^{(p)}$
and their uncertainties
determined in~\cite{dirdet-ellis}
\begin{equation}
f_{Tu}^{(p)} = 0.020 \pm 0.004, \qquad f_{Td}^{(p)} = 0.026 \pm 0.005,
\qquad f_{Ts}^{(p)} = 0.118 \pm 0.062
\end{equation}
\begin{equation}
f_{Tu}^{(n)} = 0.014 \pm 0.003, \qquad f_{Td}^{(n)} = 0.036 \pm 0.008,
\qquad f_{Ts}^{(n)} = 0.118 \pm 0.062.
\end{equation}

In this paper we calculate the quantity which is being conventionally used to compare
experimental and theoretical results  -- the cross section $\sigma_p^{SI}$
for neutralino scattering off the proton in the limit of zero momentum transfer
\begin{equation}
\sigma^{SI}=\frac{4}{\pi}{m_r^N}^2 f_N^2
\end{equation}
where $m_r^N=m_N m_{\tilde{Z_1}}/(m_N+m_{\tilde{Z_1}})$.

In our calculations we have used the CTEQ5L set of parton density 
functions~\cite{CTEQ5}
evaluated at the QCD scale $Q=\sqrt{M_{SUSY}^2-m_{\tilde{Z}_1}^2}$.

\section{Direct detection rates in the mSUGRA model}

We start with the analysis of the neutralino scattering rates in the  minimal 
supergravity model~\cite{msugra}. The mSUGRA model assumes universal boundary 
conditions at the GUT scale for scalar and gaugino masses as well as for 
trilinear $A$-parameters and therefore is defined by a set of just four 
parameters and a sign:
\begin{equation}
m_0,\ m_{1/2},\ A_0,\ \tan\beta \mbox{ and  sign}(\mu),
\end{equation}
where $\tan\beta = v_u/v_d$ parametrizes the Higgs sector. 
The top-quark pole mass is taken to be $m_t=175$~GeV.

\subsection{Experimental constraints}

Before presenting results we discuss the following experimental constraints on 
the mSUGRA model.

\begin{itemize}
\item {\bf LEP2 constraints.}
Based on negative searches for superpartners at LEP2~\cite{lep2_w1,lep2_sel}, we 
require $m_{\tw_1}>103.5$ GeV and $m_{\te_{L,R}}>99$ GeV provided $m_{\tell}-
m_{\tz_1}>10$ GeV\cite{lep2_w1}, which is the most stringent of the slepton mass 
limits.
The LEP2 experiments also set a limit on the SM Higgs boson mass: 
$m_{H_{SM}}>114.1$ GeV\cite{lep2_h}. This is of relevance since in our mSUGRA 
parameter space scans, the lightest SUSY Higgs boson $h$ is almost always SM-
like. We implement the MSSM Higgs boson mass bounds from Ref. \cite{lep2_mssmh},
although these bounds co-incide with the SM bound since in our mSUGRA
scans $m_A>120$ GeV.
\item {\bf Neutralino relic density.}
The WMAP results~\cite{wmap} set a stringent bound on the neutralino relic 
density, given by $0.0945<\Omega_{CDM}h^2<0.129$ at 95\% CL. 
The upper limit above represents a true constraint, while the 
corresponding lower
limit is flexible, since there may be additional sources of CDM such
as axions, or states associated with the hidden sector and/or extra
dimensions. 
To estimate the relic density of neutralinos in the mSUGRA model,
we use the recent calculation in Ref. \cite{bbb}. 
In that work, 
all relevant neutralino annihilation and co-annihilation
reactions are evaluated at tree level using the CompHEP\cite{comphep}
program.
The annihilation cross section times velocity is relativistically thermally
averaged\cite{graciela}, 
which is important for obtaining the correct neutralino relic 
density in the vicinity of annihilations through $s$-channel resonances.
\item{\bf $b\to s\gamma$ decay.}
The branching fraction $BF(b\to s\gamma )$ has been measured by
the BELLE\cite{belle}, CLEO\cite{cleo} and ALEPH\cite{aleph}
collaborations.  A weighted averaging of results yields $BF(b\to
s\gamma )=(3.25\pm 0.37) \times 10^{-4}$ at 95\% CL. 
To this we should add uncertainty
in the theoretical evaluation, which within the SM dominantly comes from
the scale uncertainty, and is about 10\%.
Together, these imply the bounds, 
$2.16\times 10^{-4}< BF(b\to s\gamma )< 4.34 \times 10^{-4}$. In our
study, we show contours of $BF(b\to s\gamma )$  allowing 
the reader to decide
the extent to which parameter space is  excluded~\cite{msugra-constr}. 
The calculation of $BF(b\to s\gamma )$ used
here is based upon the program of Ref. \cite{bsg}.  Our value of the SM
$b\to s\gamma$ branching fraction yields $3.4\times 10^{-4}$, with a scale
uncertainty of 10\%.
\item {\bf Muon anomalous magnetic moment.}
The muon anomalous magnetic moment $a_\mu =(g-2)_\mu/2$ has been measured to
high precision by the E821 experiment\cite{g-2}.
Comparison of the measured value against theoretical predictions gives, 
according to Hagiwara {\it et al.}\cite{HVPee}: 
$11.5<\delta a_\mu\times 10^{10}<60.7$. A different
assessment of the theoretical uncertainties\cite{HVPee} using the procedure
described in Ref.\cite{msugra-constr} gives, $-16.7< \delta a_\mu\times
10^{10}<49.1$. In view of the theoretical uncertainty,  we only present
contours of $\delta a_\mu$, as calculated using the program developed in
\cite{bbft}, and leave it to the reader to decide the extent of the
parameter region allowed by the data. 
\item {\bf $B_s\to\mu^+\mu^-$ decay.}
The branching fraction of $B_s$ to a pair of muons has been experimentally
bounded by CDF\cite{cdf}: $BF(B_s\to\mu^+\mu^- )< 2.6\times 10^{-6}$. 
While this
branching fraction is very small within the SM ($BF_{SM}(B_s \to
\mu^+\mu^-)\simeq 3.4 \times 10^{-9}$), the amplitude for the Higgs-mediated
decay of $B_s$ grows as $\tan^3\beta$ within the SUSY framework\cite{bsmm1}, 
and hence can
completely dominate the SM contribution if $\tan\beta$ is large.   In our
analysis we use the results from~\cite{bsmm}  to
delineate  the region of mSUGRA parameters excluded by the CDF upper limit on
its  branching fraction.
\end{itemize}

\subsection{mSUGRA results}

In Fig. \ref{0sig}, we show our first results of the
spin independent neutralino-nucleon elastic scattering
cross section in the $m_0\ vs.\ m_{1/2}$ parameter plane 
for $\tan\beta =10-55$, with $A_0=0$ and for $\mu <0$ 
(upper two rows) and $\mu >0$ (lower two rows).

In this figure we have applied only LEP2 constraints (black color) 
and theoretical
constraints denoted by red color which show the regions  forbidden either 
due to lack of REWSB (lower right region) or because $\tilde{\tau_1}$ is 
the LSP (upper left region).

\FIGURE[h]{
\epsfig{file=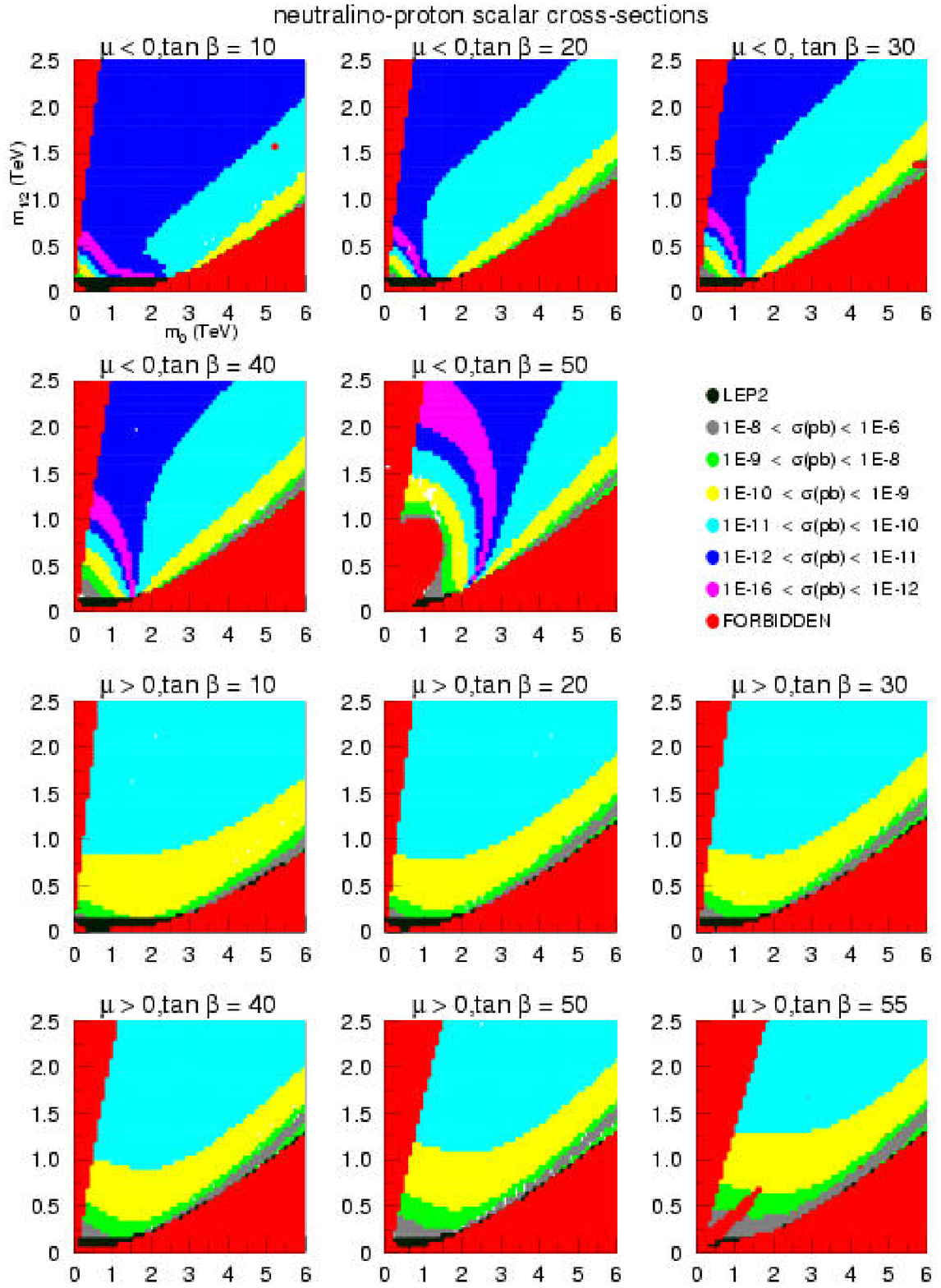,height=15cm,width=16cm} 
 \caption{\label{0sig}
 Cross section rates for spin-independent neutralino-proton  
 scattering in  the mSUGRA model for $A_0=0$.}}

One can see that for both positive and negative $\mu$ 
the the highest value for 
$\sigma_{SI}$ is about $10^{-6}$ pb. This happens in two regions: either
in the  lower left corner of the $m_0\ vs.\ m_{1/2}$ parameter space  
where the exchanged squark masses are light, 
or in the HB/FP region for large $m_0$ values and low-intermediate
$m_{1/2}$ values (gray and green colors), along  the border of the red
forbidden region where REWSB does not occur.
In this region, there is a large higgsino component to the neutralino,
which enhances the Higgs exchange contribution to $\sigma_{SI}$\cite{enhance}.
For $\mu >0$, the lowest cross sections reach below $10^{-10}$ pb and 
occur at very large values of 
$m_0$ and $m_{1/2}$ where squark masses are high, thus
suppressing the overall scattering cross section. 
For $\mu <0$, the lowest
cross section values can dip far below $10^{-12}$ pb in the magenta-shaded
regions. 
The nature of this behavior in $\sigma_{SI}$
is due to a cancellation between processes involving $up$- and $down$-type quarks
where the leading contribution comes from $t$-channel 
Higgs boson exchange~\cite{dirdet-ellis}.

\FIGURE[h]{
\epsfig{file=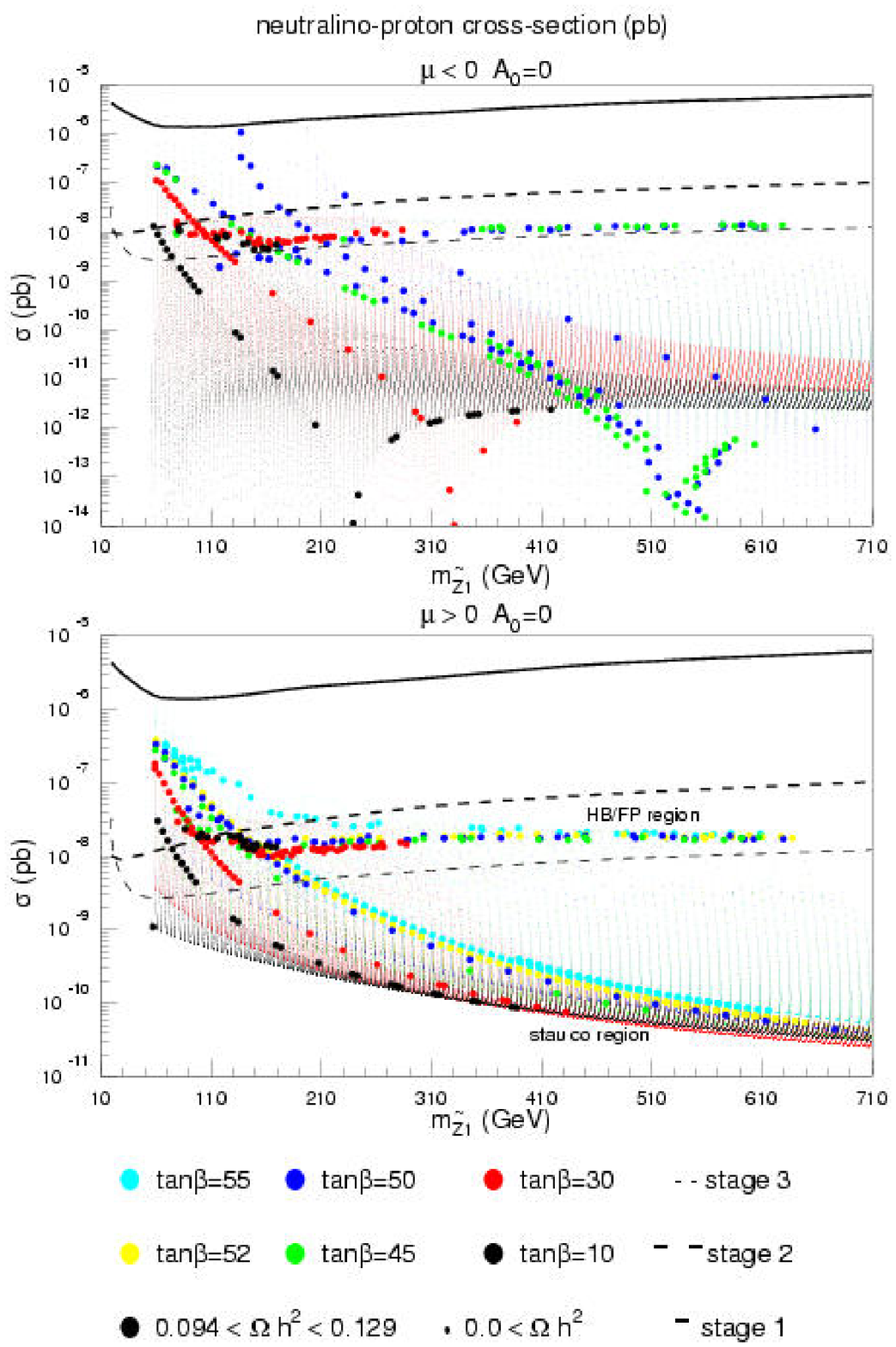,height=19.5cm}
\vspace*{-0.5cm}
 \caption{\label{0sig-mz1}
$\sigma_{SI}$ versus neutralino mass}}

In Fig.~\ref{0sig-mz1} we plot $\sigma_{SI}$ versus neutralino mass
for a random scan over mSUGRA parameter space for various $\tan\beta$ 
and $A_0=0$, for $\mu <0$ (top frame) and
$\mu >0$ bottom frame).
Parameter space points which satisfies only
theoretical and LEP2 constraints ({\it i.e.} parameter space presented in 
Fig.~\ref{0sig}) are represented by small dots.
If instead we restrict $\Omega_{\tz_1} h^2$
to be within the experimental limits $0.0945 <\Omega_{CDM}h^2<0.129$, 
only the solid circles survive. The different colors of the dots
correspond to different values of $\tan\beta$.
We can see now how restrictive  the relic density
constraint is for the parameter space and  for the range
of $\sigma_{SI}$. One can see that the maximum value  $\sigma_{SI}$
can be reduced almost by an order of magnitude for some values of $\tan\beta$.
We also show in these plots the reach of Stage 1, Stage 2 and Stage 3 
experiments. Evidently Stage 1 experiments are just now starting to explore
regions of mSUGRA model parameter space\cite{bottino}. Stage 2 and Stage 3 detectors
can explore much larger regions of parameter space, but none of the planned
detectors will be able to completely rule out SUSY dark matter.

The regions of parameter space allowed by the relic density constraint
form branches of distinct patterns.
One sort of  branch is the HB/FP region. It is exhibited by  
region  of almost constant $\sigma_{SI}\sim 10^{-8}$~pb
which is nearly independent of $\tan\beta$.
In this region, it is possible to satisfy not only the relic density
constraints, but also (as shown below) 
$a_\mu$ and $BF(b\to s\gamma )$ constraints.
One can solve the SUSY flavor and CP problems, and possibly maintain
naturalness. The Stage 3 detectors ought to have sufficient reach to
cover this interesting region of model parameter space!
It is important to note, however, that in the HB/FP the relic density can be quite low,
so that the assumed local density of dark matter should be rescaled\cite{rescale}.
Hence, in all subsequent figures we will plot $f\cdot \sigma_{SI}$, where
$f=\Omega_{\tz_1}h^2/0.094$ for $\Omega_{\tz_1}h^2<0.094$, and
$f=1$ for $\Omega_{\tz_1}h^2 >0.094$. Our reach plots will also be
determined in terms of $f\cdot \sigma_{SI}$ rather than in terms of $\sigma_{SI}$. 

In Fig. \ref{sug_ps}{\it a}.), we show $f\cdot\sigma_{SI}\ vs.\ m_{\tz_1}$ as 
a function of the complete model parameter space for $\mu <0$. 
The ranges of parameters
are listed above the figure, and include a scan over all $A_0$ values.
We also show the reach of Stage 1, 2 and 3 detectors. Most points in 
mSUGRA model parameter space are now excluded by the WMAP constraint,
so we show points with $0.094< \Omega_{\tz_1}h^2 <0.129$ by green dots,
while blue dots denote points with $ \Omega_{\tz_1}h^2 <0.094$, which would
require in addition some form of non-MSSM cold dark matter.
The corresponding plot of $f\cdot\sigma_{SI}\ vs.\ \Omega_{\tz_1}h^2$ is
shown in frame {\it b}.). The parameter space with $\mu >0$
is shown in frames {\it c.}) and {\it d.}). In the case of $\mu <0$, 
much lower cross-sections are allowed, due to the above mentioned
cancellations in quark-neutralino effective couplings. However, it is the 
$\mu <0$ sign which is more restricted by $a_\mu$ and 
$BF(b\to s\gamma )$ constraints.

\FIGURE[t]{
\epsfig{file=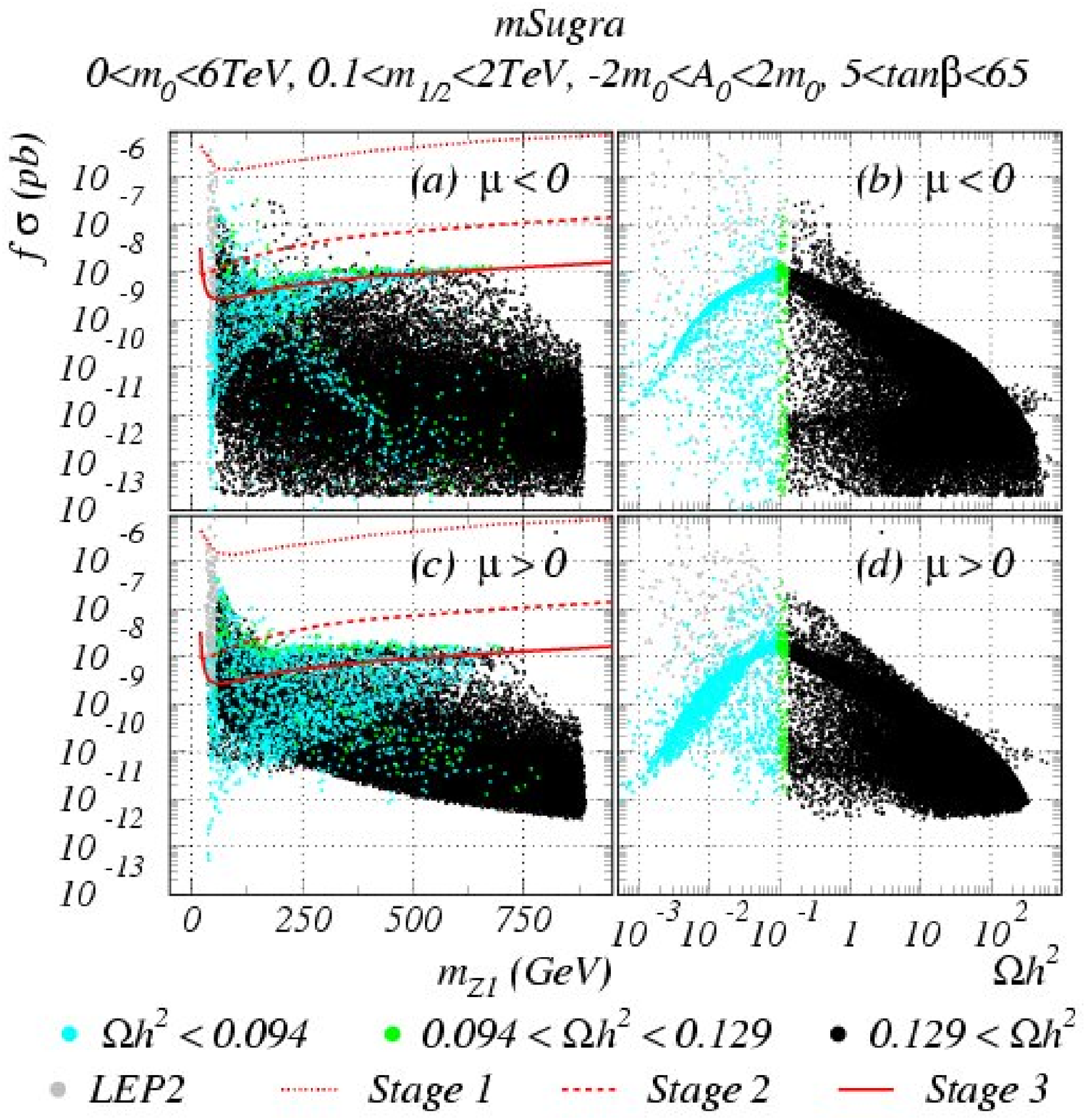,height=15cm} 
 \caption{\label{sug_ps}
In {\it a}) we plot rescaled cross section rates $vs.$ 
$m_{\tz_1}$ for a scan over mSUGRA 
parameter space with $\mu <0$. In {\it b}), we show cross section $vs.$ 
$\Omega_{\tz_1}h^2$ for the same $\mu <0$ parameter space.
In {\it c}) and {\it d}), we show the same plots, except for $\mu >0$.}}

\FIGURE[h]{
\vspace*{-0.5cm}
\mbox{
\hspace*{-1cm}
\epsfig{file=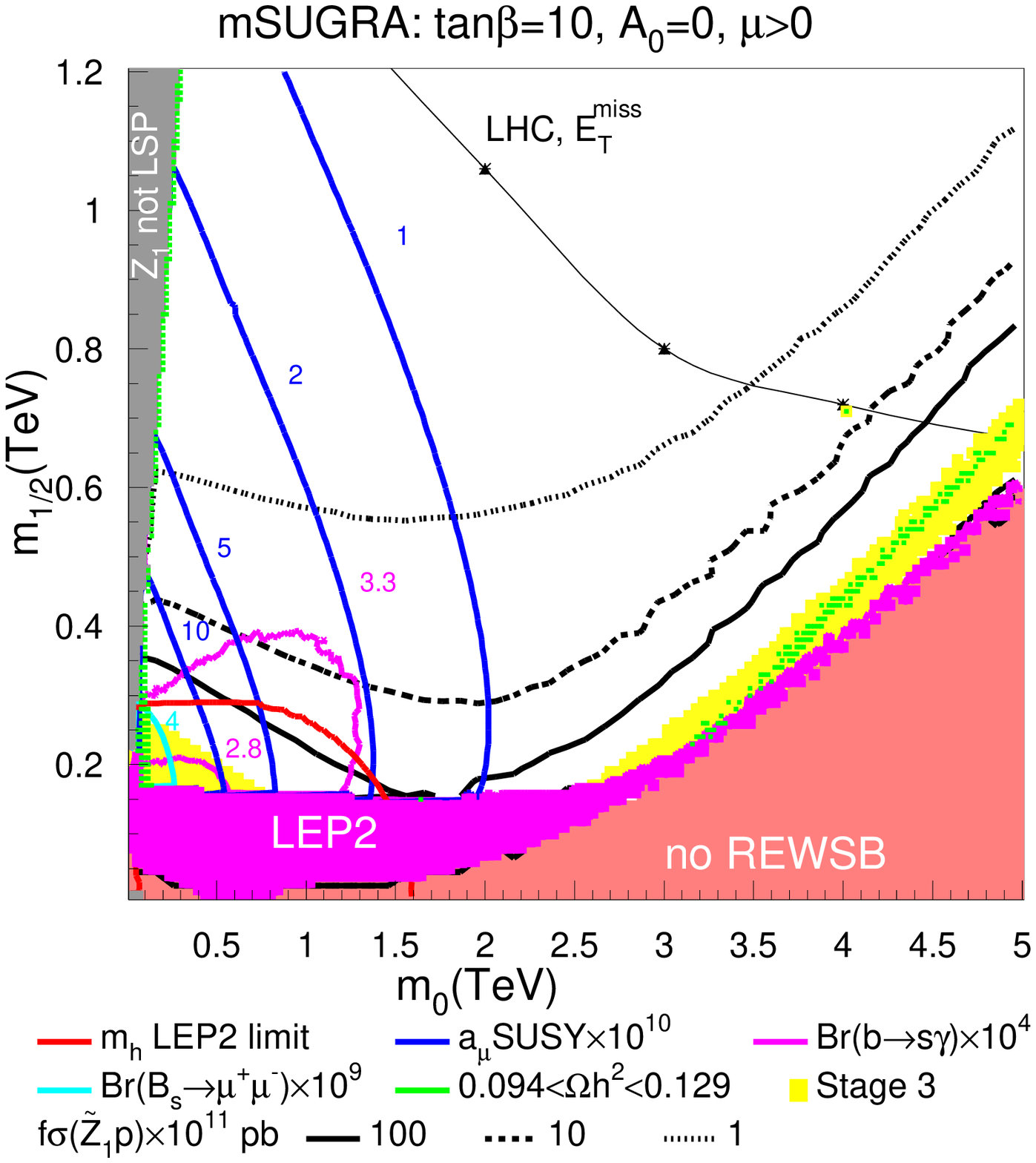,height=9.8cm}
\hspace*{-0.5cm}
\epsfig{file=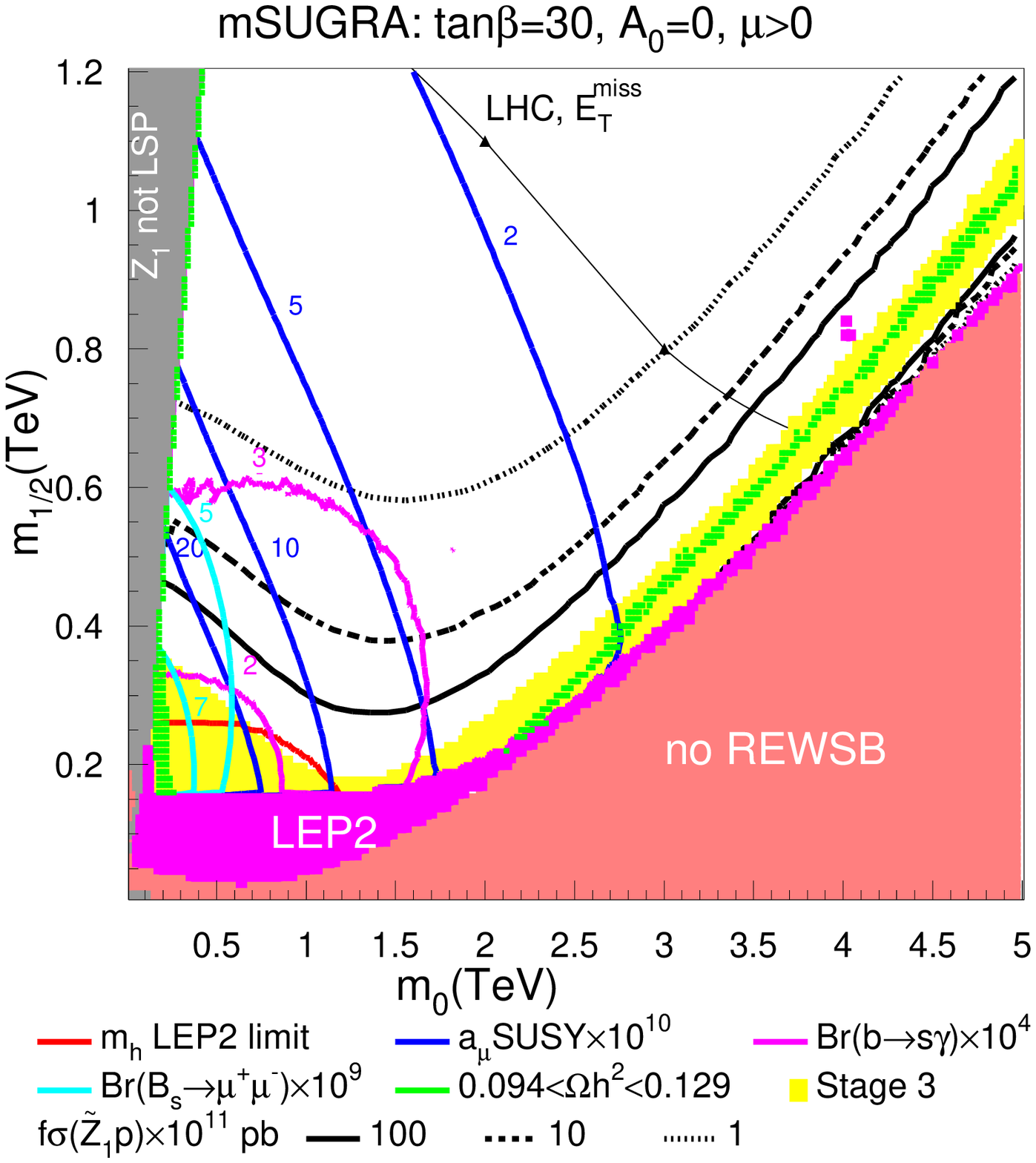,height=9.8cm}
\hspace*{-1cm}
}
\vspace*{-0.4cm}
\mbox{
\hspace*{-1cm}
\epsfig{file=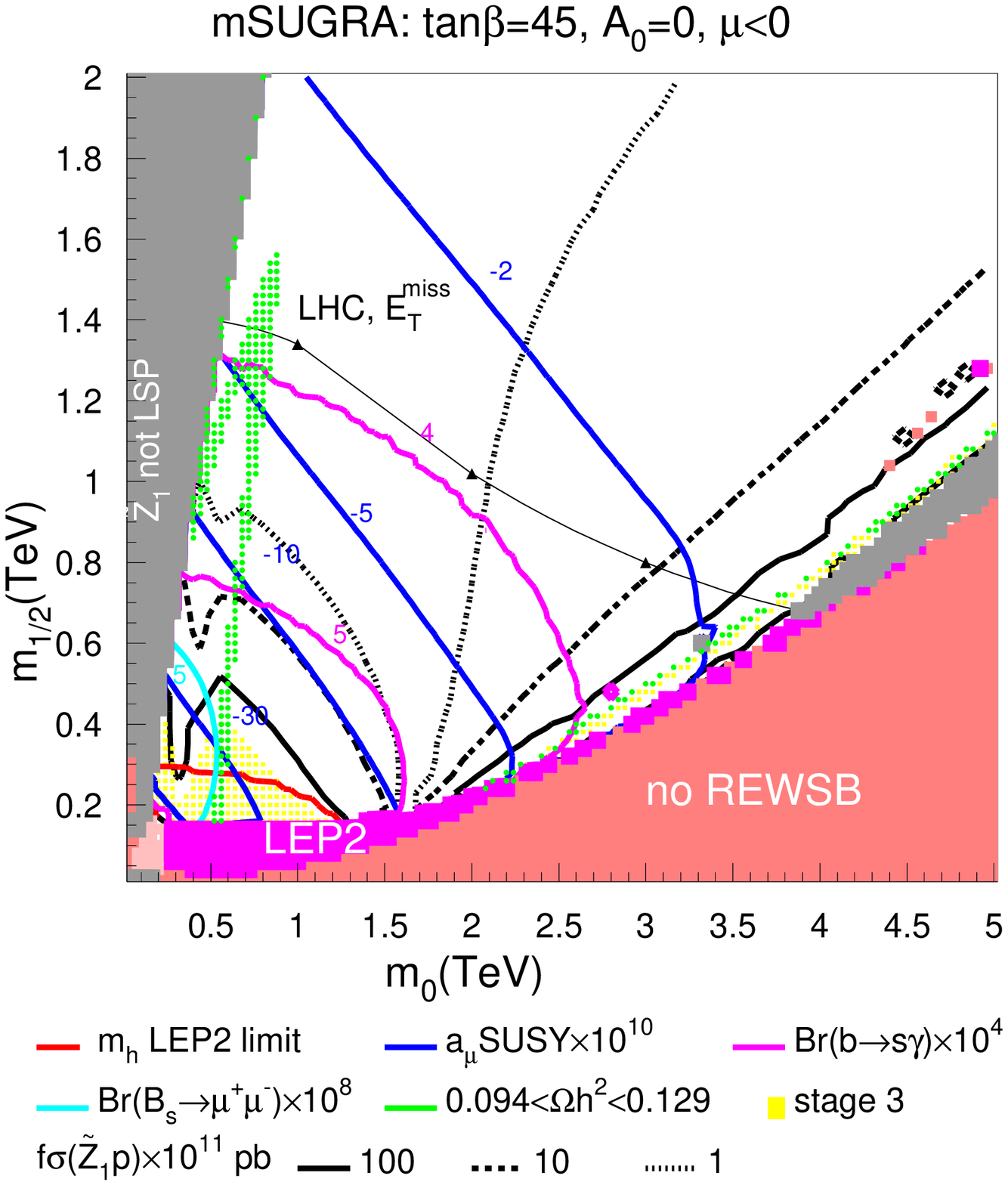,height=9.8cm}
\hspace*{-0.5cm}
\epsfig{file=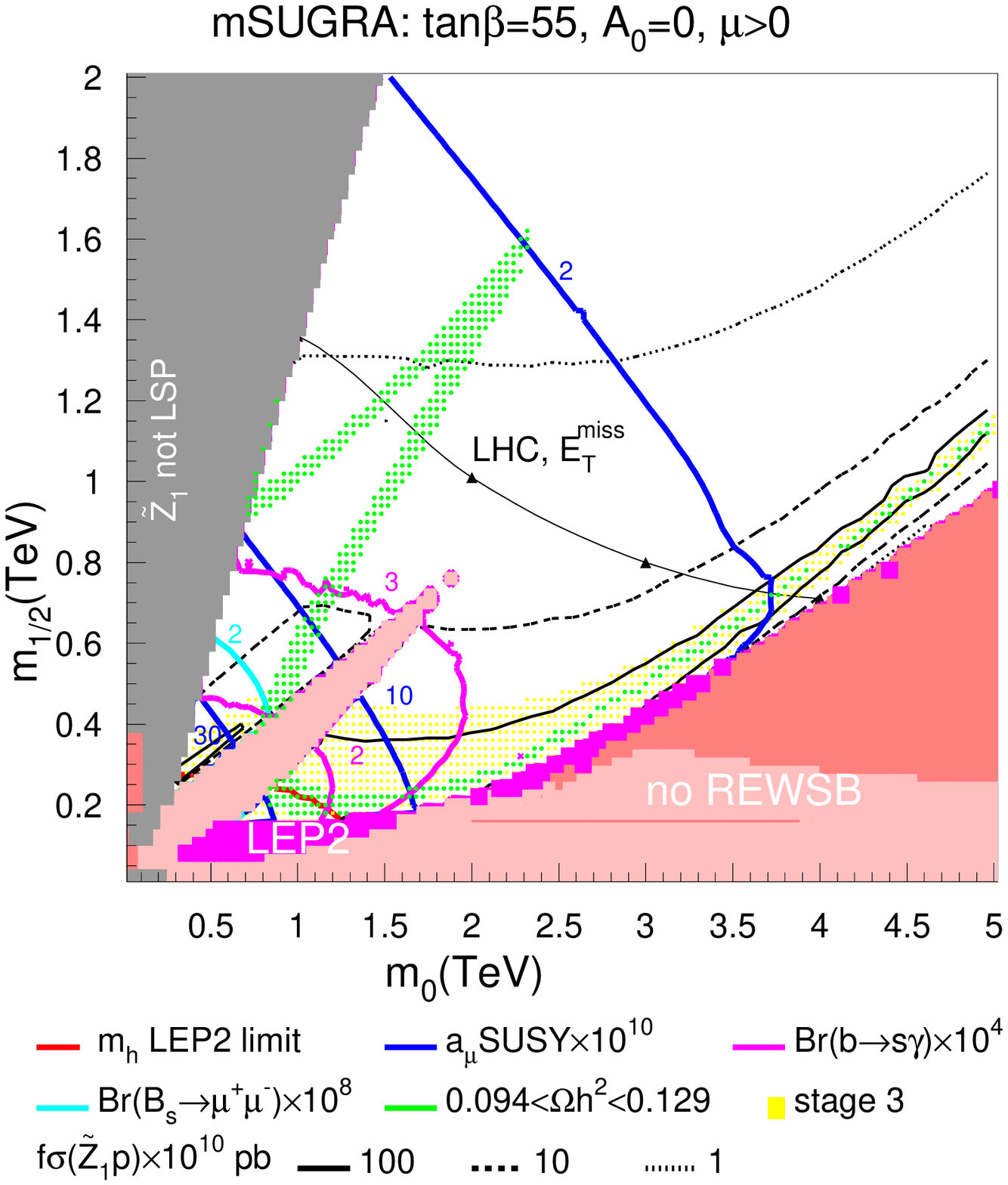,height=9.8cm}
\hspace*{-1cm}
}
\caption{\label{msugra2d}
Constraints and $\sigma_{SI}$ rates 
for mSUGRA model in $m_0$ $vs.$ $m_{1/2}$
plane:
$\tan\beta=10, \mu>0$(upper left),
$\tan\beta=30, \mu>0$(upper right),
$\tan\beta=45, \mu<0$(bottom left)
and
$\tan\beta=55, \mu>0$(bottom right),$A_0=0$.
}
\vspace*{-0.5cm}
}

Finally we present our results on $f\cdot\sigma_{SI}$
for mSUGRA  together with the above mentioned direct and 
indirect constraints
in Fig.~\ref{msugra2d} for the $m_0$ $vs.$ $m_{1/2}$ plane. 
As an example we have chosen
$\tan\beta=10,\ \mu>0$ (upper left),
$\tan\beta=30,\ \mu>0$ (upper right),
$\tan\beta=45,\ \mu<0$ (bottom left)
and
$\tan\beta=55,\ \mu>0$ (bottom right)
cases for  $A_0=0$. 
The red and black shaded regions are excluded 
due to lack of REWSB or a stau LSP respectively. 
The magenta region is excluded by LEP2 searches for 
charginos and sleptons.
The region below the red curve is excluded 
by LEP2 Higgs searches.
One can see how $BF( b\to s\gamma )$ (magenta contours), $a_\mu$ 
(blue contours),
$B_s\to\mu^+\mu^-$ (light blue contours) can further significantly
reduce MSSM parameter space and the range of $f\cdot\sigma_{SI}$.

For $\tan\beta=10$, there are three regions
of allowed relic density (which could be also recognized  by looking
at different branches of Fig.~\ref{0sig-mz1}):
\begin{itemize}
\item[1)]
lower left ``bulk'' region where neutralinos annihilate 
dominantly through $t$-channel slepton exchange;
\item[2)]
the region of $\tilde{\tau}_1-\tilde{Z}_1$ co-annihilation
which  is along the stau LSP region\cite{efos}. One should notice that this
region has highly fine-tuned relic density since a small variation in $m_0$
would lead to a large change in
$\Omega h^2$~\cite{ellis-ft,bbb};
\item[3)]
HB/FP region where $\mu$ becomes small and a large higgsino component 
of $\tilde{Z}_1$ allows for efficient annihilation into 
$WW$, $ZZ$, $Zh$ and $hh$ pairs~\cite{focus,bbb}.
\end{itemize}

In the first bulk region, $\sigma_{SI}$
is the highest and reaches  $10^{-6}-10^{-7}$~pb level~(Fig.\ref{0sig-mz1}).
It can  be easily accessible by future experiments 
on direct CDM searches as indicated by yellow region in Fig.~\ref{msugra2d}.
Unfortunately it is already excluded by the LEP2 bound on $m_h$.
It also has a rather low value of $BF(b\to s\gamma )$.

The narrow $\tilde{\tau}_1-\tilde{Z}_1$ co-annihilation 
region lacks theoretical attraction  due to large
fine-tuning in the relic density, and it is also not attractive 
for direct DM search experiments.
The typical level of $\sigma_{SI}$ is 
$10^{-10}-10^{-11}$~pb~(dashed and dotted black contours for $\tan\beta=30$
in Fig.~\ref{msugra2d}), and in this region it will be not accessible even 
by Stage 3 detectors.

Finally, from both a theoretical and experimental point of view, the
HB/FP region looks very attractive.
One can see  from Fig.~\ref{0sig-mz1} (dashed line)
and Fig.~\ref{msugra2d} (yellow shaded region)
that Stage 3 experiments on direct CDM search
can indeed cover the portion of the HB/FP region which is within the WMAP
bounds.

If we increase $\tan\beta$ to 30, as shown in Fig.~\ref{msugra2d} upper-right,
then most of the bulk region of relic density is still excluded by a combination of
Higgs mass and $BF(b\to s\gamma )$ constraints. The stau co-annihilation
region remains narrow and largely beyond reach of direct DM detection
experiments, and the HB/FP region remains consistent with all constraints,
along with having a large rate of direct DM detection. 
However, by increasing $\tan\beta$ to $45$ $(\mu<0)$ 
(Fig.~\ref{msugra2d}, bottom-left) and $\tan\beta=55\ (\mu>0)$ 
(Fig.~\ref{msugra2d}, bottom-right)
a qualitatively new region (fourth kind of region) of allowed relic density appears.
One can see the diagonal strip running from the lower-left to the upper-right 
corner. In this region, neutralinos annihilate efficiently through the 
$s$-channel $A$ and $H$ bosons which have very large widths due to large $b$- and 
$\tau$-Yukawa couplings. Though these regions are also theoretically attractive,
one can see that they will be inaccessible by neutralino direct search experiments.

Finally, we would like to stress the high degree of complementarity
between the LHC SUSY search and the direct DM search 
in restricting supersymmetric parameter space\cite{baer-brhlik,ellis}.
A very intriguing feature of the direct DM search experiments
is that they can cover most of the HB/FP region 
while LHC can explore $m_{1/2}$ only up to about 700 GeV
(largely independent of $\tan\beta$)~\cite{howie-tadas}.
On the other hand, LHC is able to completely cover
the region of neutralino annihilation through
heavy Higgs resonances and almost all of the stau co-annihilation region~\cite{howie-tadas}.
Thus, a combination of the LHC and 
Stage 3 direct DM experiments together
can cover almost the entire mSUGRA parameter space!

\section{Results for Yukawa unified models}
In this Section we present results for MSSM models with non-universal
boundary conditions motivated by $SO(10)$ SUSY GUT scenarios.
SUSY GUTs based on $SO(10)$ are theoretically very attractive.
Along with gauge coupling uification, 
they unify all matter of a single generation into a single
16 dimensional spinorial multiplet of $SO(10)$.
The 16 dimensional spinorial multiplet
contains a right-handed neutrino which becomes a SM gauge singlet. It can
be used to provide for massive neutrinos via
the see-saw mechanism. 
The massive neutrino states may play an important role
in baryogenesis via intermediate scale
leptogenesis~\cite{leptogen} due to the structure of the neutrino sector.
Finally, $SO(10)$ is an anomaly-free
group and 
explains the cancellation of triangle anomalies within the SM.

\FIGURE[h]{
\vspace*{-1cm}
\epsfig{file=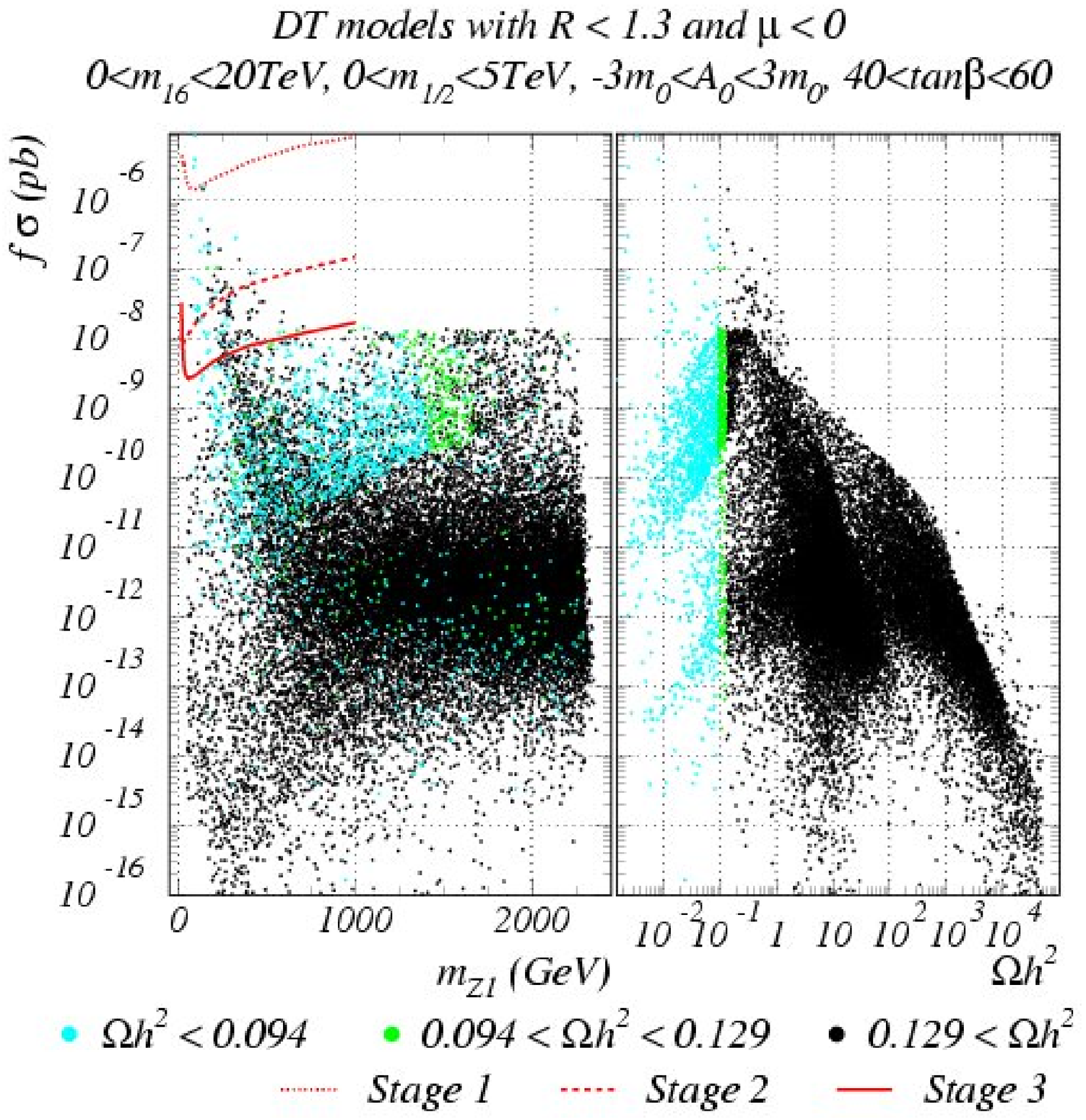,height=9.5cm}%
\vspace*{1cm}
\epsfig{file=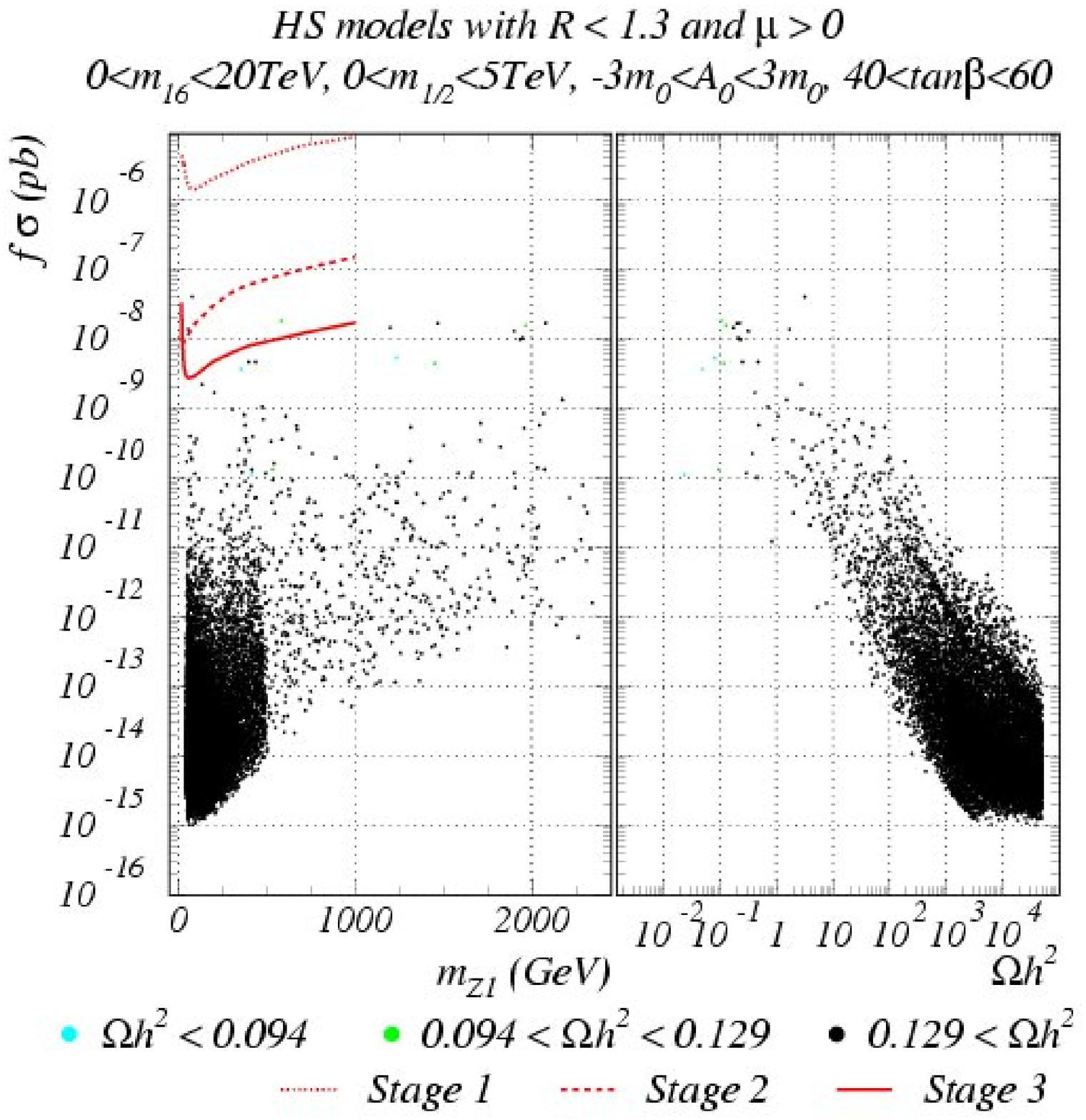,height=9.5cm} 
 \caption{\label{fig1:so10}
 Scattering  spin-independent neutralino-proton rates
 versus neutralino mass(left) and $\Omega h^2$(right)
 for $DT$(top)  and $HS$(bottom) models}
 }

In the simplest models of $SO(10)$ SUSY GUTs, the two MSSM
Higgs doublets are both present in a single 10-dimensional
representation. This structure of the Higgs sector leads to a
crucial feature of the model  --- it predicts Yukawa coupling
unification since the superpotential
contains the term:
$ \hat{f}\ni  
f\hat{\psi}({\bf 16})^T\hat{\psi} ({\bf 16})\hat{\phi} ({\bf 10})  
+\cdots .  
$
In Ref.~\cite{so10-4}, we have searched model parameter space for
solutions with a high degree of third generation Yukawa
coupling unification,
for both positive and negative values of the superpotential $\mu$ 
parameter. 
In the case of $\mu<0$, there exist regions of
SUSY parameter  space satisfying all experimental constraints and possessing
perfect Yukawa coupling unification.  
Yukawa unified solutions for $\mu>0$ were also found. These latter solutions 
required multi-TeV values of the GUT scale scalar mass parameter $m_{16}$, 
and were in general difficult to reconcile with bounds on the
relic density of neutralinos.

We have adopted results from~\cite{so10-4} using Isajet v7.64
and studied the  neutralino-proton scattering cross section
for cases with a high degree of Yukawa coupling unification.
The measure of Yukawa unification used is given by  
\begin{equation}  
R=\max(f_t,f_b,f_\tau)/\min(f_t,f_b,f_\tau),   
\label{Eq:DefR}  
\end{equation} 
 where $f_t$, $f_b$ and $f_\tau$ are the $t$, $b$ and $\tau$ Yukawa  
couplings, and $R$ is measured at $Q=M_{GUT}$.
In $SO(10)$ models, the reduction in rank of the gauge symmetry group
can lead to additional $D$-term contributions
to scalar particle masses.
In the case of 
$SO(10)\to SU(5)\times U(1)_X\to SU(3)_c\times SU(2)_L\times U(1)_Y$,
the $D$-term contributions lead to the following 
$GUT$ scale scalar mass splittings ($DT$ model)\cite{dterm}:
\begin{eqnarray*}  
m_Q^2=m_E^2=m_U^2=m_{16}^2+M_D^2 , \\  
m_D^2=m_L^2=m_{16}^2-3M_D^2 , \\  
m_N^2 = m_{16}^2+5M_D^2,\\  
m_{H_{u,d}}^2=m_{10}^2\mp 2M_D^2 ,  
\end{eqnarray*}  
where $M_D^2$ parameterizes the magnitude of the $D$-terms.  
Owing to our ignorance of the gauge symmetry breaking mechanism,  
$M_D^2$ can be taken as a free parameter,   
with either positive or negative values.
$|M_D|$ is expected to be of order the weak scale.  
An alternative scenario is for mass splittings to occur only
for GUT scale scalar Higgs  masses\cite{raby} ($HS$ model) {\it i.e.}
\begin{eqnarray*}  
m_Q^2=m_E^2=m_U^2=m_{16}^2 , \\  
m_D^2=m_L^2=m_{16}^2 , \\  
m_N^2 = m_{16}^2,\\  
m_{H_{u,d}}^2=m_{10}^2\mp 2M_D^2 .  
\end{eqnarray*}  
The $DT$ model is slightly preferred for models with $\mu<0$,
while the $HS$ model is preferred for $\mu >0$.
Both the $DT$ and $HS$ models  are  characterized by the following free   
parameters,  
\begin{eqnarray*}  
m_{16},\ m_{10},\ M_D^2,\ m_{1/2},\ A_0,\ \tan\beta ,\ {\rm sign}(\mu ).  
\end{eqnarray*}  
 We scan these models over the following parameter range:  
\begin{eqnarray}  
0&<&m_{16}<20\ {\rm TeV}, \nonumber\\  
0&<&m_{10}<30\ {\rm TeV}, \nonumber\\  
0&<&m_{1/2}<5\ {\rm TeV}, \nonumber\\  
-(m_{10}/\sqrt{2})^2&<&M_D^2<+(m_{10}/\sqrt{2})^2,\label{rangesI}\\  
40&<&\tan\beta <60,       \nonumber\\  
-3m_{16}&<& A_0<3m_{16}.\nonumber  
\end{eqnarray}  

Our results for Yukawa unified models are presented in Fig.~\ref{fig1:so10}
for $\mu <0$ $DT$ model (upper frames) and for $\mu >0$ $HS$ model
(lower frames). In each, we require $R<1.3$.
The left-most frames show $f\cdot\sigma_{SI}\ vs.\ m_{\tz_1}$, while the
right-most frames show $f\cdot\sigma_{SI}\ vs.\ \Omega_{\tz_1}h^2$.
Green points fall within the WMAP limits on the relic density, while blue 
points have $\Omega_{\tz_1}h^2 <0.094$. Black points have 
$\Omega_{\tz_1}h^2 >0.129$. We see immediately from the relatively infrequent
number of green points how restrictive the WMAP
bounds are on Yukawa unified models. Nevertheless, some Yukawa unified models
are consistent with WMAP. These models tend to have relatively heavy 
sparticle mass spectra, and usually fulfill the $\Omega_{\tz_1}h^2$
bound by having $2 m_{\tz_1}\sim m_A$ so that resonance neutralino
anihilation via heavy and pseudoscalar Higgs bosons 
occurs at a high rate\cite{so10-4}.
The WMAP allowed models tend to have direct detection rates below the reach 
of even the Stage 3 DM search experiments. We do note, however, that 
several points do occur with allowed relic densities that are accessible to
direct DM searches.

In the lower two frames, we show the corresponding results for
$HS$ models with $\mu>0$. In this case, $m_{16}$ values of $8-20$ TeV
are favored. The SUSY particle spectra is characterized by an inverted 
scalar mass hierarchy, so that the models may still satisfy 
naturalness criteria, since third generation scalar can be
relatively light. However, the very heavy spectra leads usually
to too high of values of $\Omega_{\tz_1}h^2$, and also suppresses
the direct detection rate to below the reach of even Stage 3 detectors.
Again we note that a few points do survive the relic density
constraint, and one is even within reach of Stage 3 experiments.

\FIGURE[h]{
\epsfig{file=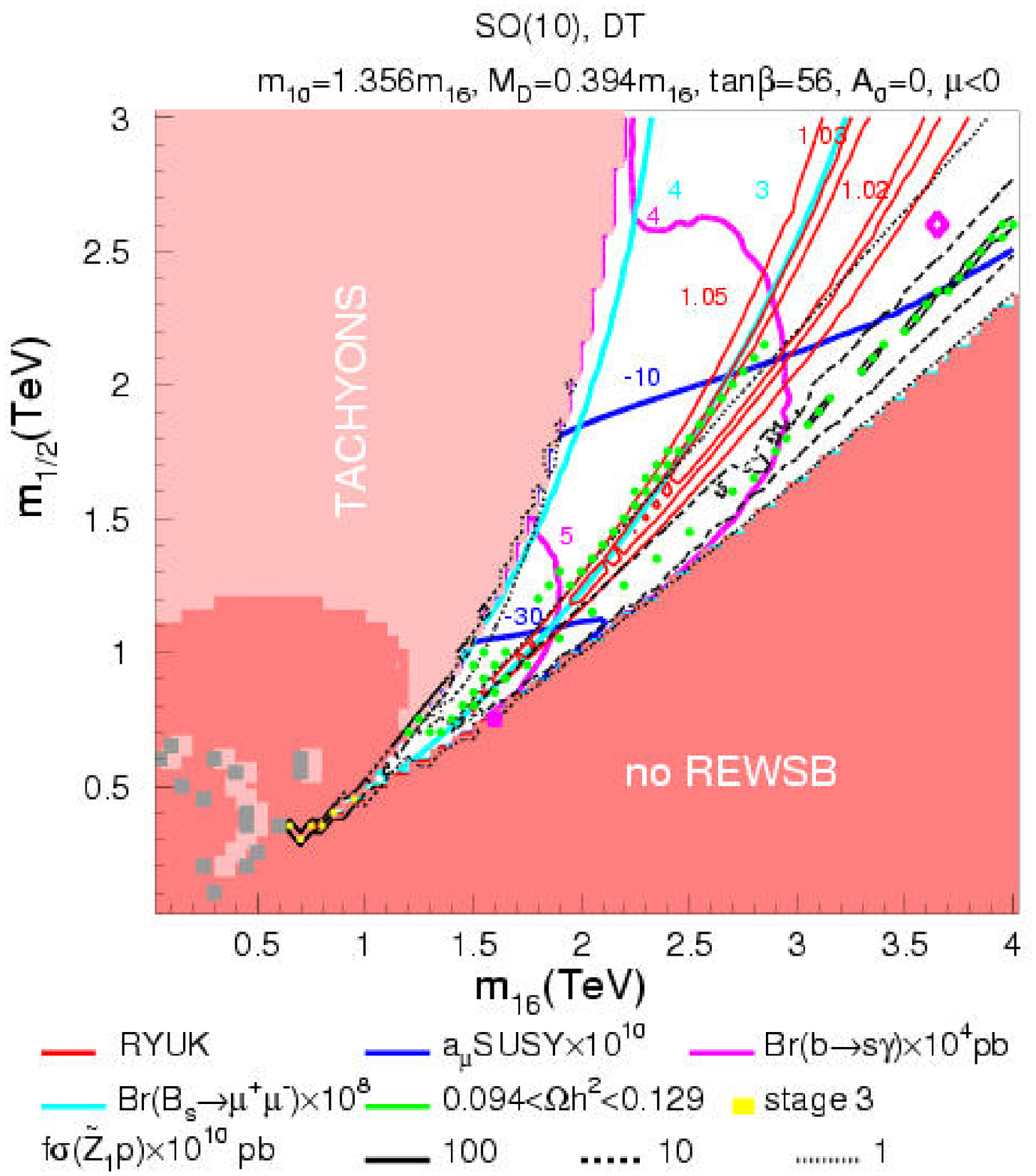,width=8cm}%
\hspace*{-0.5cm}\epsfig{file=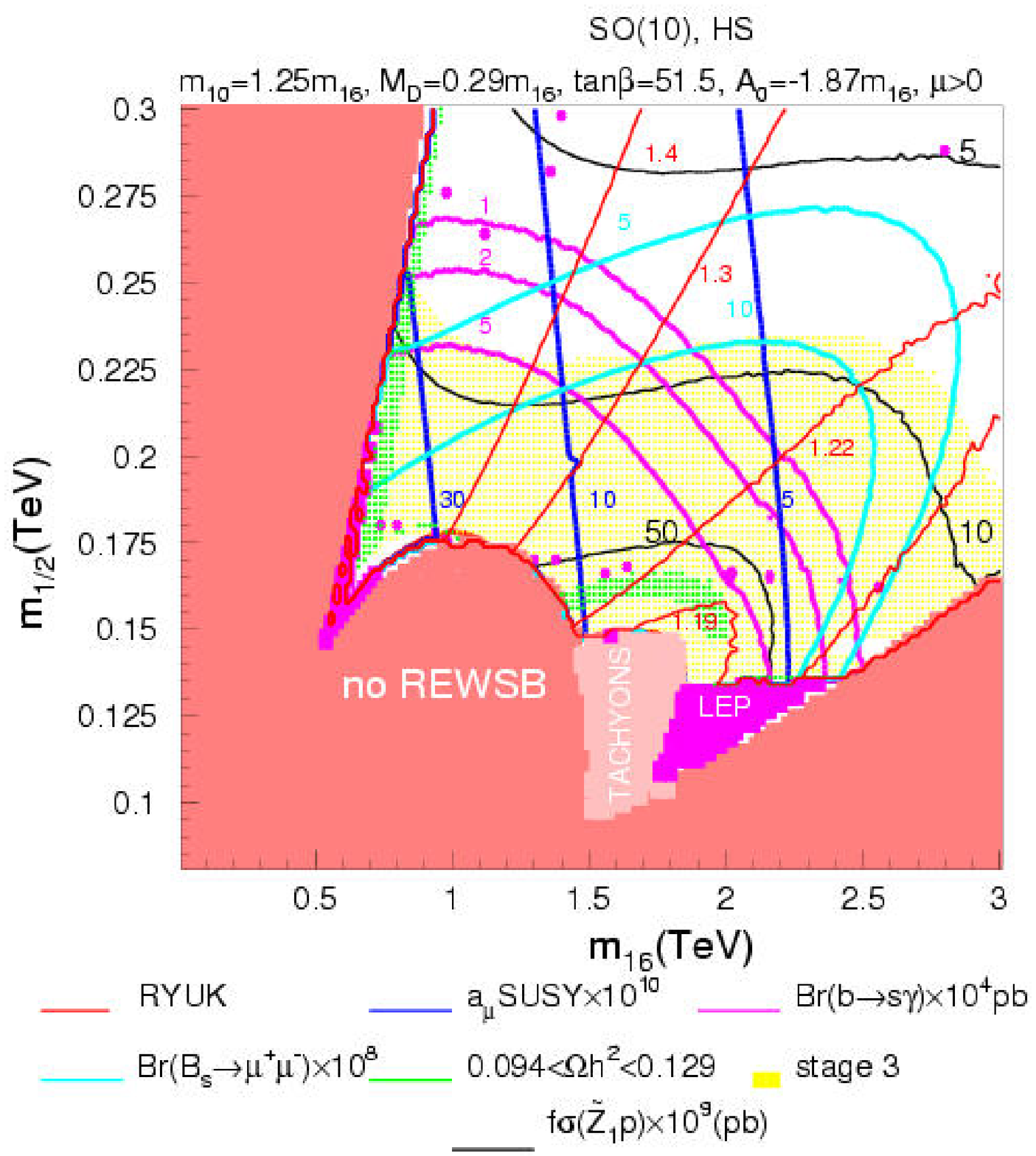,width=8cm} 
 \caption{\label{fig2:so10}
 Constraints and $\sigma_{SI}$ rates 
 in $m_0$ $vs.$ $m_{1/2}$
plane for $DT$(left) and $HS$(right) models
}
}

It is also useful to examine the Yukawa unified models
in the context of constraints from
$BF( b\to s \gamma )$, $a_\mu$ and $BF(B_s\to \mu^+\mu^- )$.
We illustrate this in Fig.~\ref{fig2:so10}, where we 
present contours of $f\cdot\sigma_{SI}$ 
together with experimental constraints in the 
$m_{16} \ vs. \ m_{1/2}$ plane.
The left frame of Fig.~\ref{fig2:so10} presents results
for  $\mu<0$ and  
$m_{10}=1.356 m_{16}$, $M_D=0.394 m_{16}$, $A_0=0$ and $\tan\beta=56$,
which leads to a central region of the plot with good Yukawa unification
(illustrated by the $R<1.02-1.05$ contours.
One can see that the region of good Yukawa coupling
unification can satisfy all experimental constraints with
a small  $b\to s\gamma$ pull. The valid region is marginal  for the future
direct CDM search experiments which is indicated by the yellow region. The 
value of $f\cdot\sigma_{SI}$ can be quite hight  $\sim
10^{-7}$~pb in the lower left corner but  
$BF(b\to s \gamma)$ is   too high in this
region (magenta contour) and therefore  it is excluded. In addition
this corner of parameter space is excluded by experimental constraint
on $a_\mu$ since its  deviation from the SM is below 
$a_\mu = -30\times 10^{-10}$ 
(blue contour) level. 

We illustrate results for the $HS$ model with $\mu>0$ in the right most frame,
where the parameter set
was chosen to give reasonable values of the neutralino 
relic density  and to have Yukawa
unification at the order of 20\%.
The left most side has a good relic density value but poor Yukawa unification.
The low $m_{1/2}$ region has an area of good relic density 
(due to neutralino annihilation through the light Higgs $h$ pole) and
Yukawa coupling unification below $R=1.19$. 
While much of this region is accessible to direct DM searches,
much of it is also gives values of $BF(b\to s\gamma )$
outside of the window of expectation.

\section{Gaugino mediated SUSY breaking models with
non-universal gaugino masses}

As mentioned in the introduction, the formulation of 
SUSY GUTs in extra dimensions yields elegant solutions to many
of the problems encountered by four-dimensional models\cite{4d-prob}.
Recently, SUSY GUT models have been formulated in five and even 
higher dimensions\cite{5d+so10}.
Models have been constructed
wherein the doublet-triplet splitting problem is simply solved, and where proton decay
is either suppressed or forbidden. $R$-parity conservation may also
naturally arise. All these facets may arise without the need for 
the large Higgs 
representations which are needed to break the $GUT$ symmetry in 4-d models.

In 5-d models, compactification of the extra dimension on
an $S^1/(Z_2\times Z'_2)$ orbifold leads to two inequivalent fixed points
identified as separated 4-dimensional branes within the 5-d bulk.
Only fields with positive parities under the $Z_2$ and $Z_2'$ symmetries
have massless Kaluza-Klein modes. The $Z_2$ parity assignments are chosen so that
the $N=1$ SUSY in 5-d, which normally reduces to $N=2$ SUSY in 4-d, in fact
breaks to $N=1$ SUSY in 4-d. 
Superfields containing MSSM matter are assumed to exist on the observable brane,
while the ``hidden'' brane is set up to accomodate SUSY breaking.
The $Z_2'$ parity assignments are made so that
the grand unified symmetry is broken on the hidden brane.
The set up is well suited to accomodate SUSY breaking via 
gaugino mediation, 
wherein SUSY breaking is communicated from the hidden brane to the visible brane
via gauge superfields which propagate in the bulk\cite{inoMSB}. In gaugino mediated SUSY
breaking (inoMSB), the scalar, trilinear and bilinear soft SUSY breaking
masses are loop suppressed, and can effectively be taken to be zero. 
The gaugino masses
however are non-zero. 

Several intriguing scenarios arise in gaugino mediation. In all cases,
in the 4-d effective theory below $min(M_c,M_{GUT})$, 
$m_0\sim A_0\sim 0$, while non-zero $\mu$ and $B$ can be generated.\footnote{
Analyses of models with non-universal gaugino masses and non-zero scalar masses
and $A$-terms are presented in Ref. \cite{dbranemodels}.}
These latter parameters are traded for $\tan\beta $ and $M_Z$, as is usual
where electroweak symmetry is broken radiatively.
The scenarios are given as follows.
\begin{itemize}
\item 1. If $SU(5)$ is the gauge symmetry on the hidden brane, 
then universal gaugino masses will result. In this case, gaugino mediation
leads to a stau LSP unless additional above-the-GUT-scale RGE running
is allowed (Schmaltz-Skiba case)\cite{ss,bdqt}.
\item 2. If $SU(5)$ is the gauge symmetry of the bulk, but is broken to
the SM gauge symmetry on the hidden brane, then independent gaugino masses
are induced. The effective 4-d theory on the visible brane is just the
MSSM, with parameter space $M_1(M_c),\ M_2(M_c),\ M_3(M_c),\ \tan\beta$
and $sign(\mu )$.
\item 3. If $SO(10)$ is the bulk symmetry, and is broken to
$SO(6)\times SO(4)$ (isomorphic to the Pati-Salam group 
$SU(4)\times SU(2)_L\times SU(2)_R$) on the hidden brane,
and $SO(10)$ is broken to $SU(5)$ on the visible brane via the usual
Higgs mechansim, 
then the visible sector model obeys parameter space is given by
$M_2(M_c),\ M_3(M_c),\ \tan\beta$ and $sign(\mu )$, where
$M_1={2\over 5}M_3+{3\over 5}M_2$ at $Q=M_c$ 
(Dermisek-Mafi case)\cite{pati-salam,radovan-paper}.
\item 4. If $SO(10)$ is the bulk gauge symmetry, and flipped 
$SU(5)'\times U(1)'$
is the hidden brane symmetry, then the parameter
space is given by  $M_1(M_c),\ M_2(M_c),\ \tan\beta$ and $sign(\mu )$,
where $M_3(M_c)=M_2(M_c)$ (Barr-Dorsner case)\cite{flipped-su5}.
\end{itemize} 
In this section, we explore the direct dark matter detection rates for
the latter three cases of non-universal gaugino mediation.
In these models, the additional parameter freedom arising from
non-universal gaugino mediation can be exploited to solve
the slepton LSP problem, instead of above-the-GUT scale running
suggested by Schmaltz and Skiba. Hence,
in our calculations, we adopt the choice of  $M_c=1\times 10^{16}$~GeV, {\it i.e.}
near but just below the usual unification scale of  
$M_{GUT}\simeq 2\times 10^{16}$~GeV. This choice preserves the prediction of gauge 
coupling unification.

\FIGURE[h]{
\epsfig{file=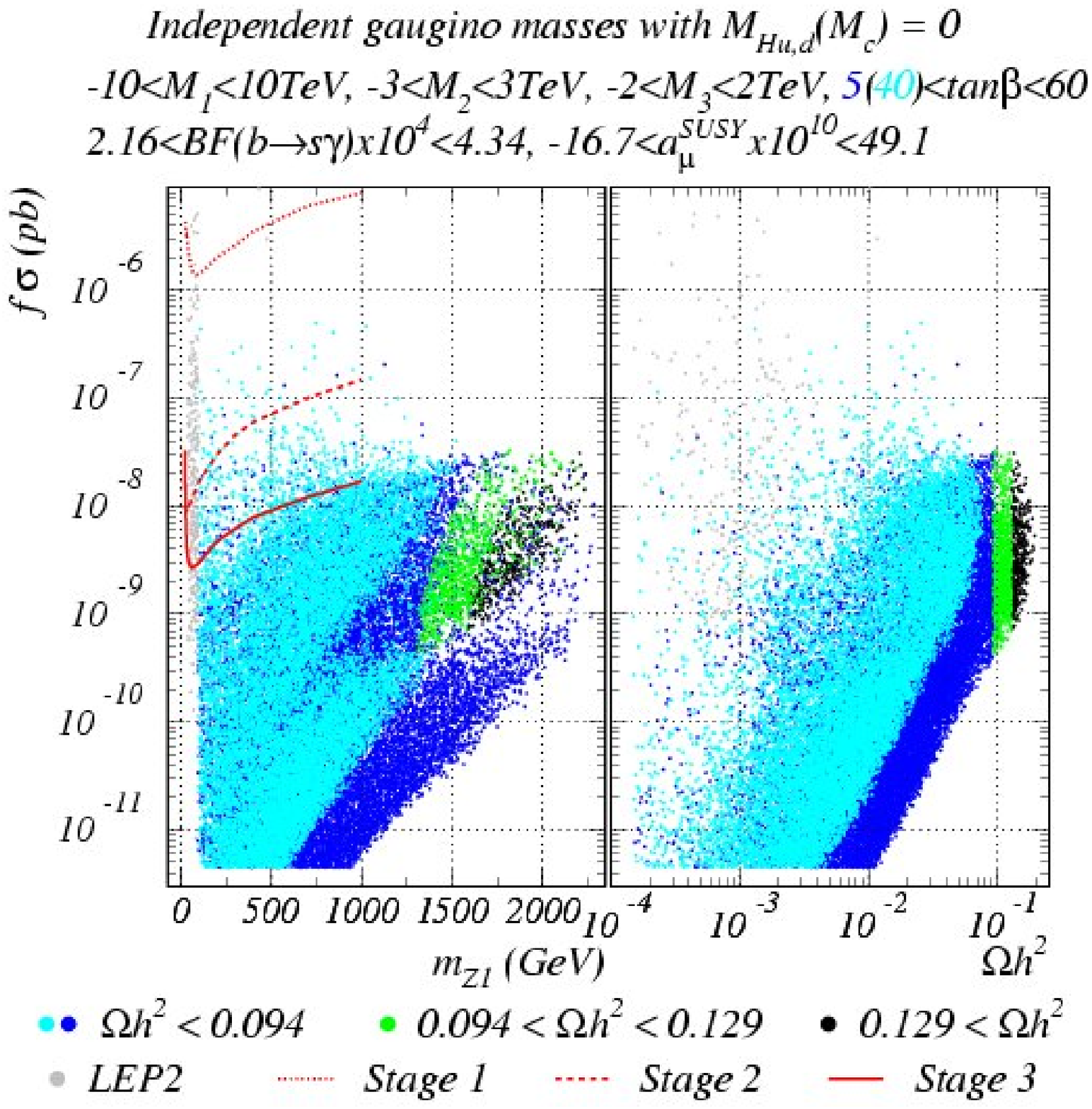,width=16cm}
 \caption{\label{fig:indep1} Constraints and $\sigma_{SI}$ rates
 for independent gaugino mass scenario.
 }}
Our main results are shown in Fig. \ref{fig:indep1} 
for case 2 with independent gaugino masses at the compactification scale.
We have scanned over the parameter space limits as listed at the top
of the figure. Solutions which yield a spectrum with a neutralino LSP
are listed by dots. These solutions usually have a lightest neutralino
with a substantial wino or higgsino component, since $M_1(M_c)$ must be
chosen to be large enough to contribute to RG running of $m_{\ttau_R}^2$ so that
it avoids becoming the LSP. 
Neutralinos with a large wino or higgsino component can annihilate
at high rates to $WW$, $ZZ$ and $Zh$ pairs so that they generally give a 
rather low value of the relic density\cite{birk}. Thus, supersymmetric dark matter would
have to be augmented by other forms of dark matter to match the constraints
on $\Omega_{CDM}h^2$ from WMAP and other measurements.
Solutions with $\Omega_{\tz_1}h^2>0.129$ are given by black dots, and
would be excluded. Green dots denote solutions with $0.094<\Omega_{\tz_1}h^2<0.129$, 
and would not require other sources of dark matter. Finally, the light and dark blue 
dots have $\Omega_{\tz_1}h^2<0.094$. The light blue solutions correspond to those
with high $\tan\beta >40$, while light blue solutions have $5< \tan\beta <40$. 
We see from the plot that a significant number of solutions are accessible
to Stage 1, Stage 2 and Stage 3 detectors. 

While the scans over the complete parameter space are comprehensive, it
can be instructive to plot results in particular parameter space planes.
We present a scan of the $M_2\ vs.\ M_3$ parameter space in 
Fig.~\ref{fig:indep2} for $\mu >0$ and $\tan\beta =10$ with $M_1=1$ TeV
(left frame) and $\tan\beta =40$, $M_1=1.5$ TeV (right frame).
It is easy to see that taking $M_2$ or $M_3$ too large will result in 
a model with a non-neutralino LSP, while taking $M_3$ too small results in
a breakdown in REWSB. Taking $M_2$ too small results in a chargino mass
in violation of LEP2 limits. The green regions give viable solutions which have
in addition $\Omega_{\tz_1}h^2<0.129$. The yellow regions in addition
should be accessible to Stage 3 direct DM detection experiments.
We also show various contours of $m_h$ values, $b\to s\gamma$ branching fraction
and $a_\mu$. Whether or not any of these additional constraints are invoked, 
it is clear that at least some of the parameter space would be accessible to direct detection 
searches.
%
\FIGURE[h]{
\hspace*{-0.5cm}\epsfig{file=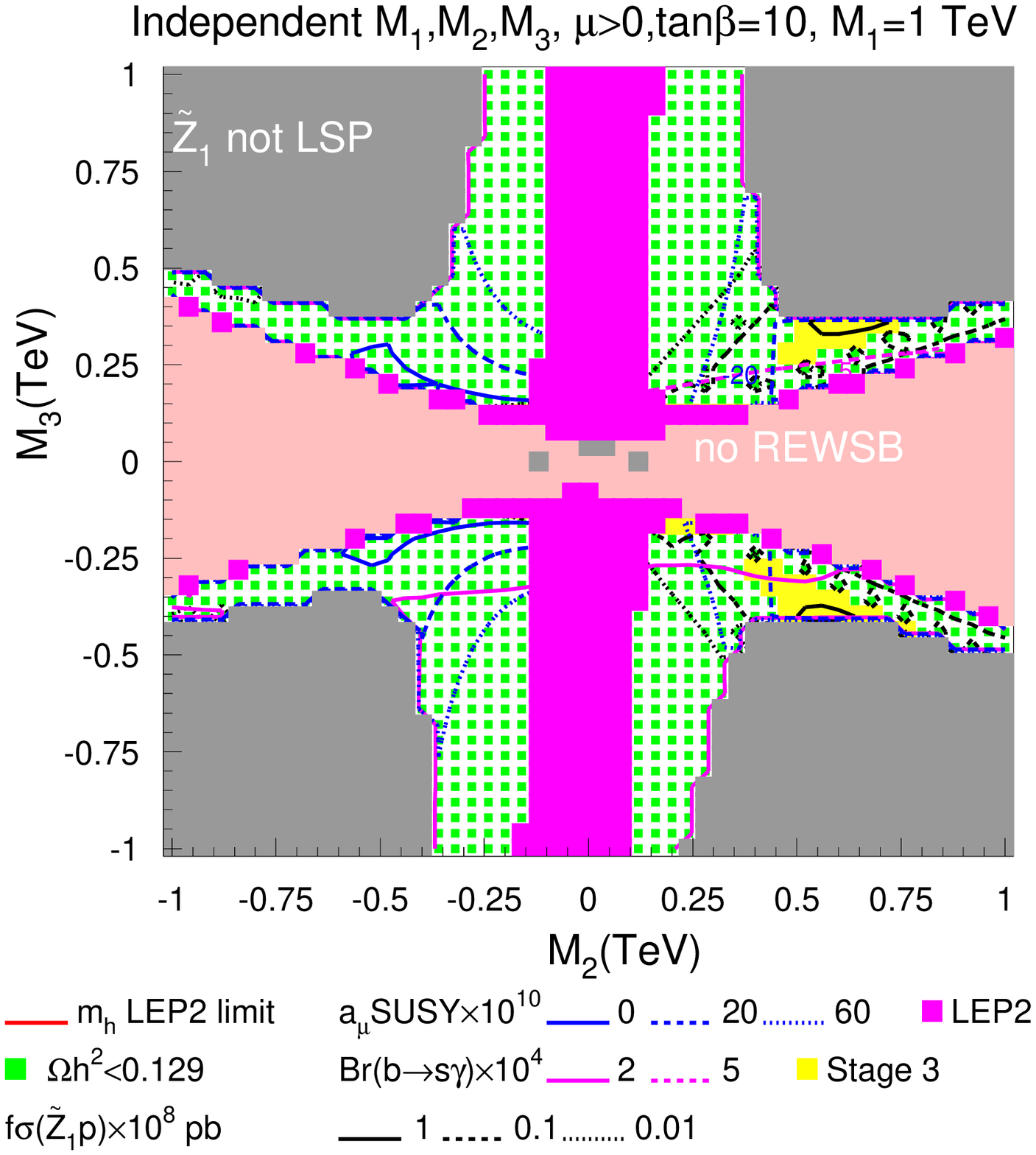,width=8.cm}%
\epsfig{file=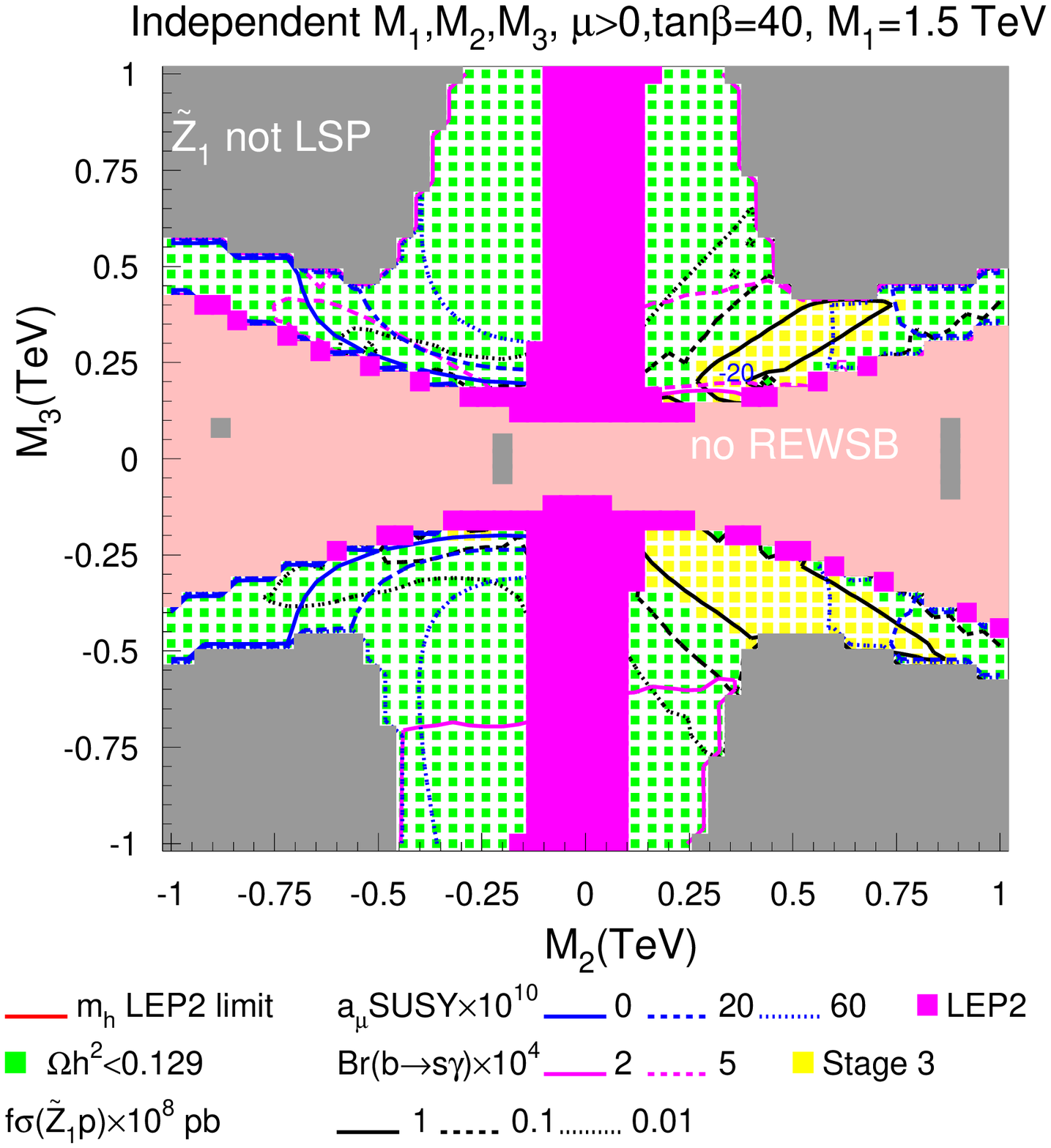,width=8.cm}
\caption{\label{fig:indep2} Constraints and $f\cdot\sigma_{SI}$ rates
 for independent gaugino mass scenario in $(M_2 - M_3)$ plane
 }}

If case 3 with $SO(10)$ breaking to the Pati-Salam group is invoked, then the
parameter space becomes more restrictive, since now $M_1$ is determined by the
choice of $M_2$ and $M_3$. 
In this case, we again plot parameter space in Fig. \ref{fig:pati2} 
in the $M_2\ vs.\ M_3$ plane
for $\tan\beta =5$ (left) and $10$ (right), with $\mu >0$.
\FIGURE[h]{
\hspace*{-0.5cm}\epsfig{file=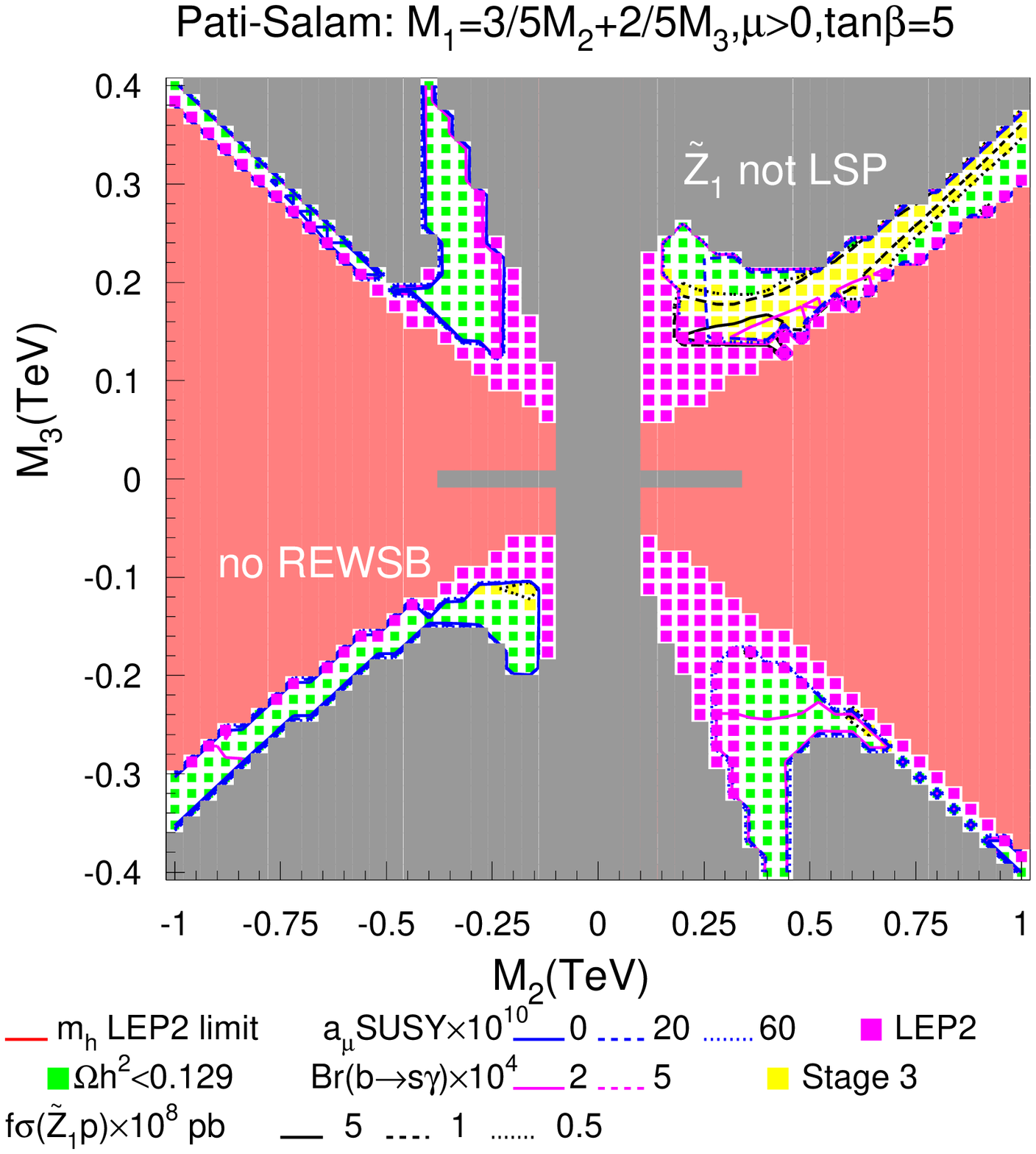,width=8.cm}%
\epsfig{file=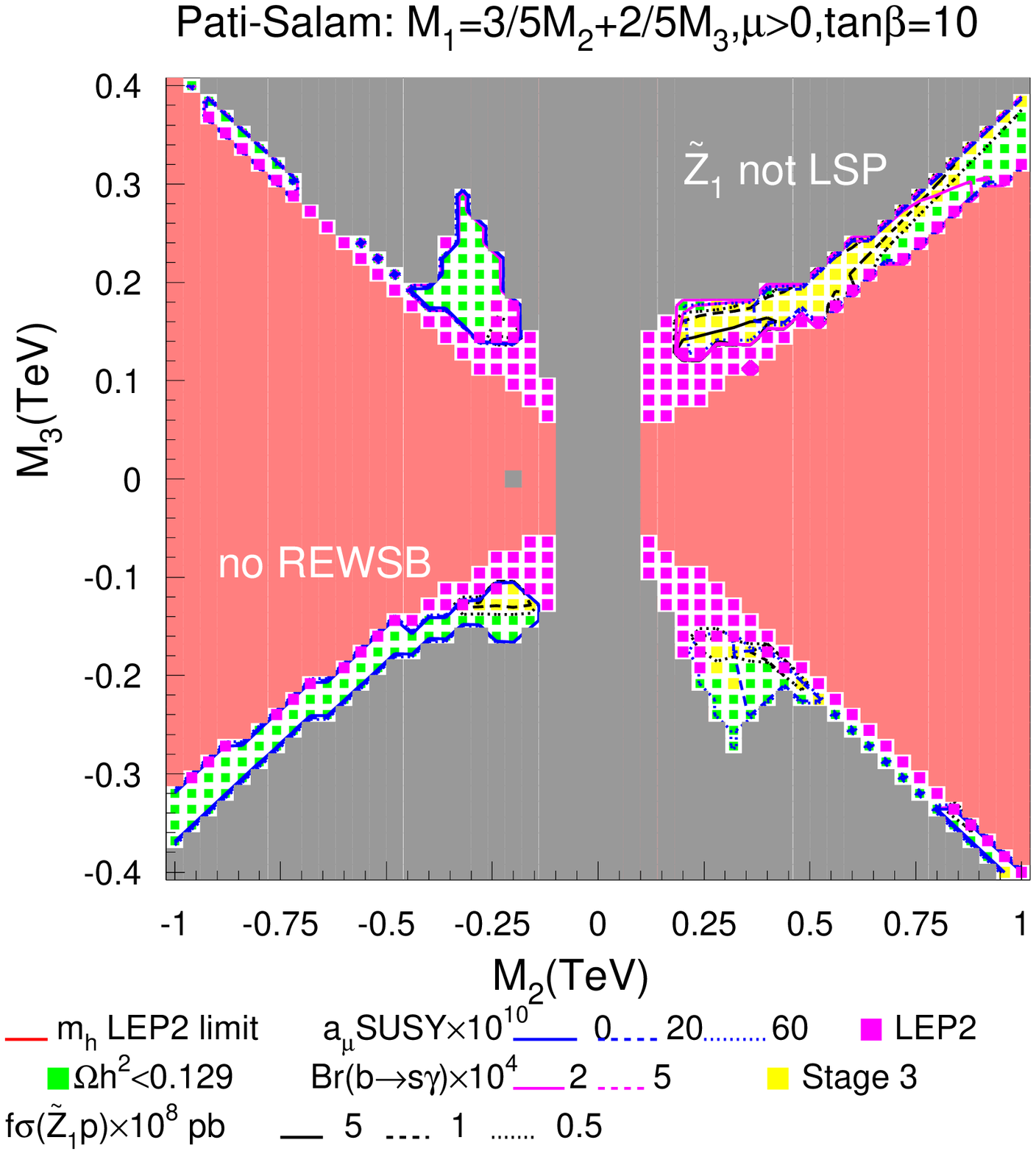,width=8.cm}
 \caption{\label{fig:pati2} Constraints and $f\cdot\sigma_{SI}$ rates
 for  Pati-Salam scenario in $(M_2-M_3)$ plane
 }
}
The viable parameter space is again denoted by the green color, with
experimental constraints indicated by various contours.
Much higher values of $\tan\beta$ result in almost no viable parameter 
space\cite{radovan-paper}. 
The yellow regions again denote parameter space accessible to
Stage 3 experiments.

The $M_2\ vs.\ M_1$ parameter space with $M_2=M_3$ for
the case of $SO(10)$ breaking to flipped $SU(5)$ is
presented  in Fig.~\ref{fig:flip1}, for $\tan\beta =10$ (left) and 40 (right), 
and for $\mu >0$.
\FIGURE[h]{
\hspace*{-0.5cm}\epsfig{file=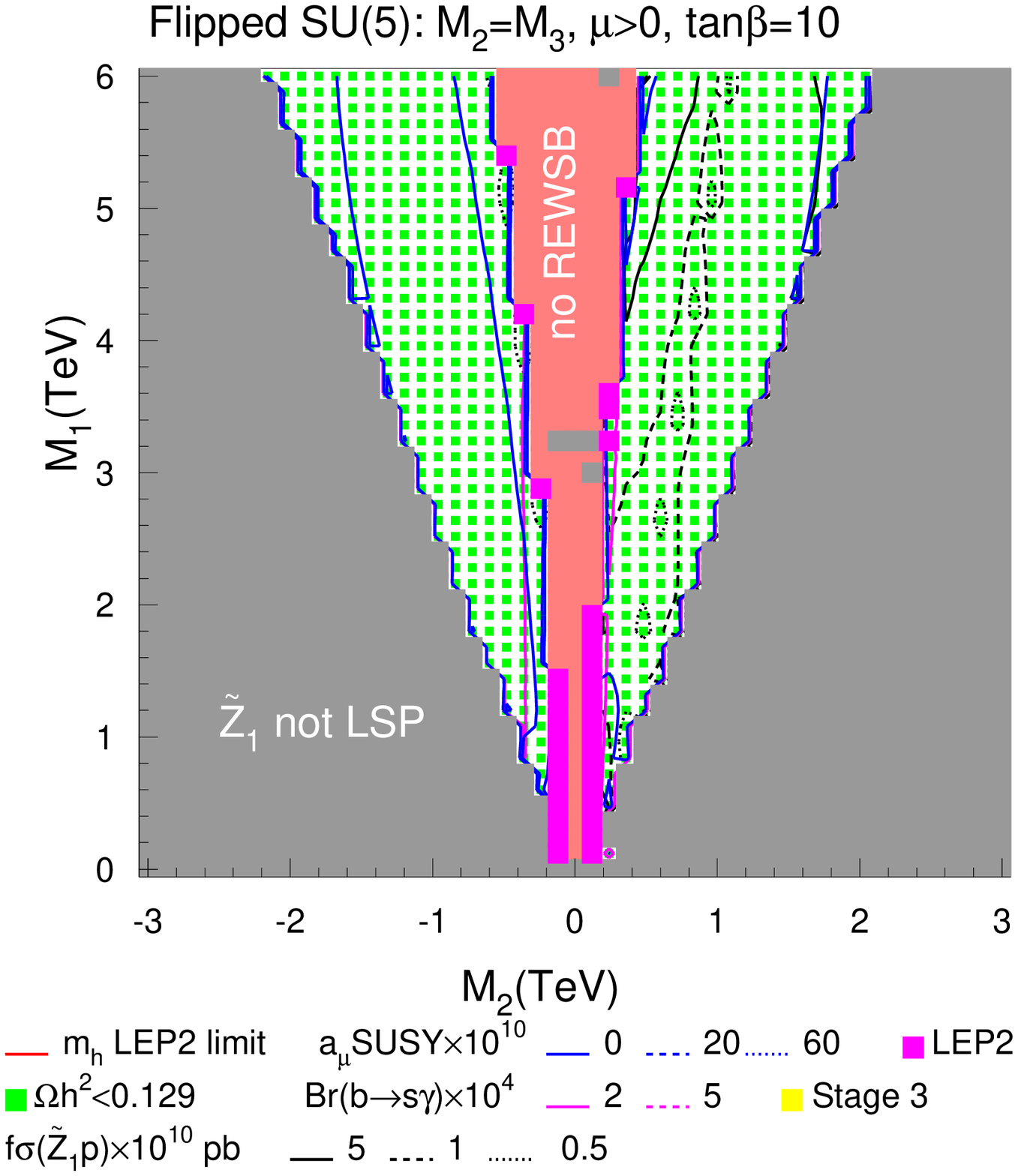,width=8.cm}%
\epsfig{file=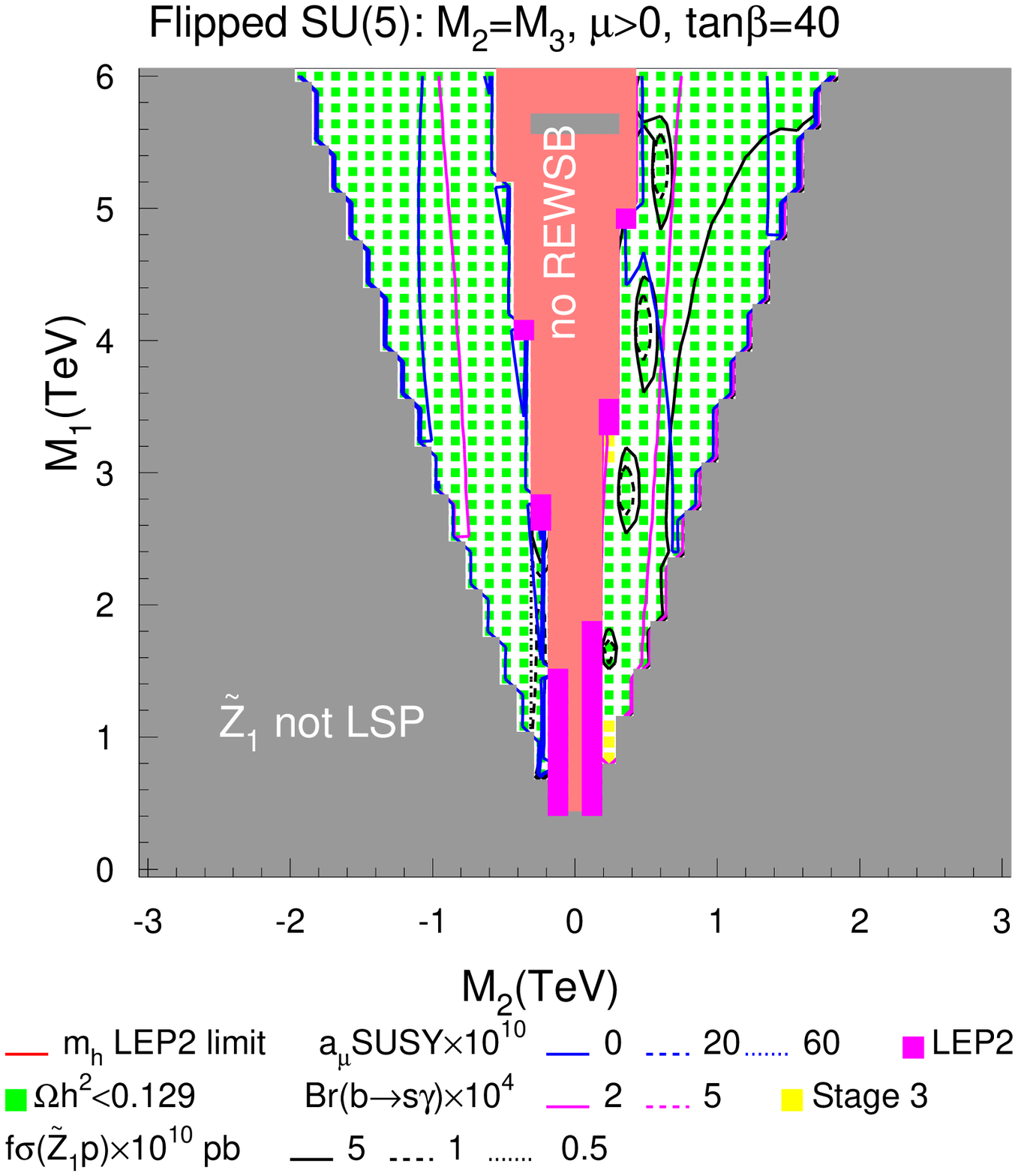,width=8.cm}
 \caption{\label{fig:flip1} Constraints and $f\cdot\sigma_{SI}$ rates
 for  flipped $SU(5)$ model in $(M_2-M_3)$ plane
 }}
 One can see that the parameter space opens up with increasing $M_1$, and
moreover, that the relic density is safely below the WMAP upper limit.
The region with very low values of $M_2$ is excluded because the chargino
mass falls below limits from LEP2. For higher (but still low) values of $M_2$,
the lightest neutralino is dominantly wino-like, and has a large
scattering cross section with protons. However, the low value of relic density
causes $f\cdot \sigma_{SI}$ to fall below the level accessible to Stage 3
direct detection searches.

\section{Conclusions}
\label{sec:conclude}

In this paper, we have evaluated the neutralino-proton scattering 
cross section in 1.) the mSUGRA model, 2.) in $SO(10)$
SUSY GUT models with Yukawa coupling unification, and 3.) in extra-dimensional
SUSY GUT models with gaugino-mediated SUSY breaking and
non-universal gaugino masses. Our results for the mSUGRA model give
a comprehensive scan of $f\cdot \sigma_{SI}$ over model parameter space using
Isajet v7.65. We also compared the DM detection rate against other
constraints including those on the neutralino relic density from
recent WMAP data, the $b\to s\gamma$ branching fraction and the
muon anomalous magnetic moment $a_\mu$. The relic density
constraint yields only several viable regions of parameter
space: the bulk region at low $m_0$ and $m_{1/2}$, the stau
co-annihilation region at low $m_0$, the HB/FP region at large $m_0$,
and the Higgs resonance annihilation region at large $\tan\beta$.
The bulk region contains relatively light slepton masses of
300-600 GeV, so that neutralinos can annihilate via $t$-channel
graphs in the early universe. The relatively light sparticle mass spectra
associated with this region yields observable rates for
direct detection of dark matter, but can also lead to values of $m_h$, 
$BF(b\to s\gamma )$ and/or $a_\mu$ in violation with experimental
limits. The stau co-annihilation region and Higgs resonance region
typically have heavy sparticle mass spectra, and low direct detection rates,
while the HB/FP region has low $|\mu |$ values and neutralinos with a
significant higgsino component. This latter quality allows
for large neutralino annihilation cross sections in the early universe,
and also for observable rates for direct DM detection. 
Direct DM search experiments can probe most of the viable
HB/FP region. This is fortuitous, since this region is 
difficult to access via the CERN LHC or a linear $e^+e^-$ collider.
Meanwhile, the CERN LHC can probe essentially all the
Higgs annihilation region at large $\tan\beta$, and almost all
of the stau co-annihilation region. Our analysis illustrates
the complementarity of collider searches for supersymmetric matter 
with direct searches for relic DM particles.

We also examined two classes of SUSY GUT models. The first,
Yukawa unified $SO(10)$ models, have non-universal scalar masses.
The constraint of Yukawa coupling unification along with 
indirect constraints from relic density, $b\to s\gamma$ and 
$a_\mu$ point to rather heavy SUSY spectra in these models.
For $\mu <0$, the relic density constraint can be accomodated
by living in the Higgs annihilation corridor. But then sparticles
must be rather heavy, and direct detection rates are usually, but not always, 
low. For $\mu >0$, multi-TeV scalar masses are required, and the
model is characterized by a radiatively induced inverted mass
hierarchy. The very heavy scalars suppress neutralino annihilation
cross sections, so the relic density is typically very large,
and direct detection rates are very low.

The other class of models, extra dimensional SUSY GUTs with non-universal
gaugino masses, require $m_0\sim A_0\sim 0$. But then the
compactification scale gaugino mass $M_1$ must be typically larger than
$M_2$ and $M_3$ to avoid a charged LSP. In this case, the LSP
can be a neutralino, but with a significant wino or higgsino
component (in mSUGRA, the LSP is almost always bino-like).
This leads to large neutralino annihilation rates and a low 
relic density, but it also leads to large direct detection cross sections.
A portion of the parameter space of the independent gaugino mass case and the Pati-Salam
case can be probed by direct detection experiments. However, in the case of flipped $SU(5)$
gauge symmetry on the hidden brane, the relic density is so low that the 
rescaled direct detection cross sections are almost always below the level of even Stage 3 experiments.

\section*{Acknowledgments}
 
We thank M. Drees for comments on the manuscript.
JO would like to thank V. Hagopian and H. Prosper for encouraging 
his work on this topic.
This research was supported in part by the U.S. Department of Energy
under contract number DE-FG02-97ER41022.
	
%

\end{document}